\newif\ifsubmission
\submissiontrue 

\documentclass[journal]{IEEEtran}
\usepackage{makeidx}
\usepackage{graphicx}
\usepackage{psfrag}
\usepackage{verbatim}
\usepackage{amsmath}
\usepackage{url}
\usepackage{float}
\usepackage{amsmath}
\usepackage{epsfig}
\usepackage{amssymb}
\usepackage{amsfonts}
\usepackage{xcolor}
\usepackage{epstopdf}
\usepackage{subcaption}
\usepackage{array}
\usepackage{booktabs}
\usepackage[bookmarks=false]{hyperref}
\usepackage{enumitem}
\usepackage{algorithmic}
\usepackage{algorithm}

\ifsubmission
  \newcommand{\TODO}[1]{}
  
  \newcommand{\ecomment}[1]{}
  \newcommand{\acomment}[1]{}
\else 
 \newcommand{\TODO}[1]{\textcolor{red}{TODO: #1}}
  
  \newcommand{\ecomment}[1]{ {  {\color{blue} {[\bf eva: #1]} }}}
  \newcommand{\acomment}[1]{ {  {\color{red} {[\bf aza: #1]} }}}
\fi

\newcommand{\sys}{RankMap}
\newcommand{\p}{{\bf p}}
\renewcommand{\a}{{\bf a}}
\renewcommand{\r}{{\bf r}}
\newcommand{{\x}}{{\bf x}}
\newcommand{{\bhat}}{\hat{\bf{b}}}
\newcommand{{\xhat}}{\hat{\bf{x}}}
\newcommand{{\hhat}}{\hat{\bf{h}}}
\newcommand{{\zhat}}{\hat{\bf{z}}}

\newcommand{\z}{{\bf z}}

\newcommand{\y}{{\bf y}}

\renewcommand{\b}{{\bf b}}

\renewcommand{\v}{{\bf v}}

\newcommand{\D}{{\bf D}}
\newcommand{\G}{{\bf G}}
\newcommand{\E}{{\bf E}}
\newcommand{\A}{{\bf A}}

\newcommand{\U}{{\bf U}}
\newcommand{\V}{{\bf V}}

\newcommand{\R}{\mathbb{R}}

\title{\sys: A Framework for Distributed Learning from Dense Datasets}
\author{
Azalia Mirhoseini$^1$, Eva. L.  Dyer$^2$, Ebrahim. M. Songhori$^3$, Richard Baraniuk$^4$, and Farinaz Koushanfar$^5$\\
\vspace{4pt}
{Dept. of Electrical and Computer Engineering, Rice University, Houston, TX$^{1,3,4,5}$}\\
\vspace{3pt}      
{Dept. of Physical Medicine and Rehabilitation, Northwestern University$^2$}\\
 {\{azalia$^1$, ebrahim$^3$, richb$^4$, farinaz$^5$\}@rice.edu,}
 {edyer$^2$@ric.org}
} 

\begin{document}
\maketitle

\begin{abstract}
This paper introduces  \sys{}, a platform-aware end-to-end framework for efficient execution of a broad class of iterative learning algorithms for massive and dense datasets. Our framework exploits data structure to scalably factorize it into an ensemble of lower rank subspaces. The factorization creates sparse low-dimensional representations of the data, a property which is leveraged to devise effective mapping and scheduling of iterative learning algorithms on the distributed computing machines. We provide two APIs, one matrix-based and one graph-based, which facilitate automated adoption of the framework for performing several contemporary learning applications. To demonstrate the utility of \sys{}, we solve sparse recovery and power iteration problems on various real-world datasets with up to 1.8 billion non-zeros. Our evaluations are performed on Amazon EC2 and IBM iDataPlex servers using up to 244 cores. The results demonstrate up to two orders of magnitude improvements in memory usage, execution speed, and bandwidth compared with the best reported prior work, while achieving the same level of learning accuracy.

\end{abstract}

\begin{IEEEkeywords}
Dense and Big Data, Large-Scale Distributed Computing, Iterative Machine Learning, Subspace Factorization.
\end{IEEEkeywords}

\section{Introduction}
Many modern learning algorithms are based on exploring the underlying patterns, correlations, and dependencies present across the signals in the dataset. Some prominent examples of such algorithms and their applications are linear or penalized regression \cite{hoerl1970ridge}, power iterations \cite{journee2010generalized}, belief propagation \cite{conf:yedidia2000}, and expectation maximization \cite{book:montgomery2012, conf:nodelman2012}. In all of these settings, solving the underlying objective function requires iterative updates of parameters of interest until convergence is achieved. Such iterative updates often require matrix multiplications that involve the data dependency or Gram matrix. In scenarios where data is too large to fit on a single computing node and must be distributed, iterative dependency-based updates become challenging as they incur large computation and communication costs. 

To facilitate parallel computing, a number of distributed abstractions that target iterative learning algorithms have been developed, e.g., Pregel \cite{jour:malewicz2010}, Spark \cite{conf:zaharia2010}, and GraphLab \cite{jour:low2010}.  
These abstractions adopt a graph-parallel model which consists of a \textit{sparse graph} and a kernel function that runs in parallel on each vertex \cite{OSDI:Gon12}. Performance gains are achieved due to the communication-minimizing partitioning of the graph and effective control of data movement.

While graph-parallelism has been shown to accelerate machine
learning and signal processing tasks for sparse graphs,
this approach cannot be readily applied when the data exhibit
a {\em non-sparse dependency matrix}. The storage of such data in a graph format becomes very inefficient as it requires storing a large number of edges (pairwise non-zero correlation values) for each vertex (data sample). In addition, finding efficient graph cuts and partitions is infeasible when dense dependencies exist. Data with dense dependencies appear in a wide range of fields such as computer vision, medical image processing, boundary element methods and their applications, and N-body problems \cite{jour:chen1998, jour:gray2000}. Thus, finding efficient solutions for running iterative learning algorithms on densely dependent data is of paramount importance.

In this paper, we introduce \sys{}, a novel distributed framework for efficient execution of a broad class of iterative learning algorithms on datasets with dense but structured dependencies. Our key observation is that, despite the apparent high dimensionality of data, in many settings, dense datasets are low rank or lie on a union of much lower dimensional subspaces. We exploit this property to reduce the overhead associated with processing dense data dependencies---a factor which has rendered the currently available graph-parallel abstractions impractical for processing dense datasets. \sys{} provides a set of interfaces and transformations that enable efficient data-aware content analysis, as well as coordinated mapping and optimization to the specifics of the underlying hardware components. \sys{} significantly improves the runtime and energy consumption of the learning algorithms by reducing the amount of required computation, distributed system communication, and storage.

To accelerate large matrix multiplications required to compute an iterative update, we decompose dense but structured data and rewrite it as a product of two matrices with far fewer non-zeros than the original data. The decomposed data is then used in subsequent iterative learning algorithms in lieu of the original dense data. We introduce a host of automated methods for partitioning the decomposed factors and ordering the computation flow in a distributed setting. The partitioning algorithm is efficient (within a bound from the optimum) and has a constant runtime. We introduce two different representations and accompanying computational models (a matrix-based and a vertex-centric model) to compute an update. Depending on the data domain and the sparsity of the decomposed components, there are different regimes where each of these two models deliver highest efficiency.

We provide APIs for both matrix-based and vertex-centric iterative update models on the transformed data. Our APIs are open-source and available at \cite{rankmapapi15}. Our matrix-based API uses the general Message Passing Interface (MPI). Our vertex-centric API is based upon the GraphLab programming model. We develop an efficient mapping of the iterative computations on the sparsified decomposed data within the constraints of the GraphLab distributed framework. Both APIs are written in C++. We evaluate \sys{} on the Amazon Elastic Cloud (EC2) computing service and IBM iDataPlex computer cluster. Our experiments utilize up to 244 cores on 12 large computing nodes. 

Our explicit contributions are as follows:

\begin{itemize}[noitemsep]
\item We propose \sys{}, a large-scale learning framework that proposes sparse transformations for accelerating iterative learning algorithms on dense but structured data.
\item We introduce a scalable transformation which maps structured (low-rank) data onto two matrices which contain far fewer number of non-zeros. A systematic way to tune the transformation error to achieve a desired level of accuracy in the learning applications is provided.
\item We develop efficient distributed computational models to conduct iterative updates on the decomposed data. Highly effective partitioning methods for the decomposed data along with data-aware performance bounds are provided. 
\item We perform proof-of-concept evaluation on applications including eigenvalue decomposition, denoising, and classification that demonstrate up to two orders of magnitude improvement in runtime and memory footprint.
\end{itemize}

This paper is organized as follows. In Section \ref{sec:global}, we provide a global overview of \sys{}. In Section \ref{sec:related}, we review  related work. In Section \ref{sec:factor}, we introduce our novel data transformation algorithm. In Section \ref{sec:impact}, we study the impact of the decomposition error on learning and provide a method for automatically tuning \sys{} to produce a user-specified learning error. In Section \ref{sec:fact}, we introduce schemes for cost-efficient distributed partitioning, along with the details of our graph and matrix-based computational models. In Section \ref{sec:eval}, we provide evaluation results on multiple synthetic and real-world datasets. Finally, in Section \ref{sec:ext}, we discuss the practicality of our framework, describe domain-specific use-cases of each of the proposed computational models, and conclude.

\section{\sys{} Framework}\label{sec:global}
\subsection{Overview and approach}
In this paper, we introduce \sys{}, a distributed data-aware framework that efficiently executes learning algorithms applied to the data. The main idea underlying our approach is to leverage structure in large collections of data to decompose the correlation (Gram) matrix of the data such that the system costs (e.g, runtime, memory, and energy) associated with iterative learning algorithms are significantly reduced.

Let the data matrix $\A \in \R^{m \times n}$ denote a collection of $n$ signals of $m$-dimensions, and $\G=\A^T\A$ denote the Gram matrix. Many learning algorithms iteratively update a solution vector, denoted by $\x$, according to an update function of the following form: 
\begin{eqnarray}\label{eq:updategeneral}
\x^{iter+1}= f(\G\x^{iter}),
\end{eqnarray}
where $f( \cdot )$ is a low-complexity function and $iter$ is the current iteration. Examples are included in Section \ref{sec:targetapps}.

\begin{figure}
\centering
	\includegraphics[width=0.48\textwidth]{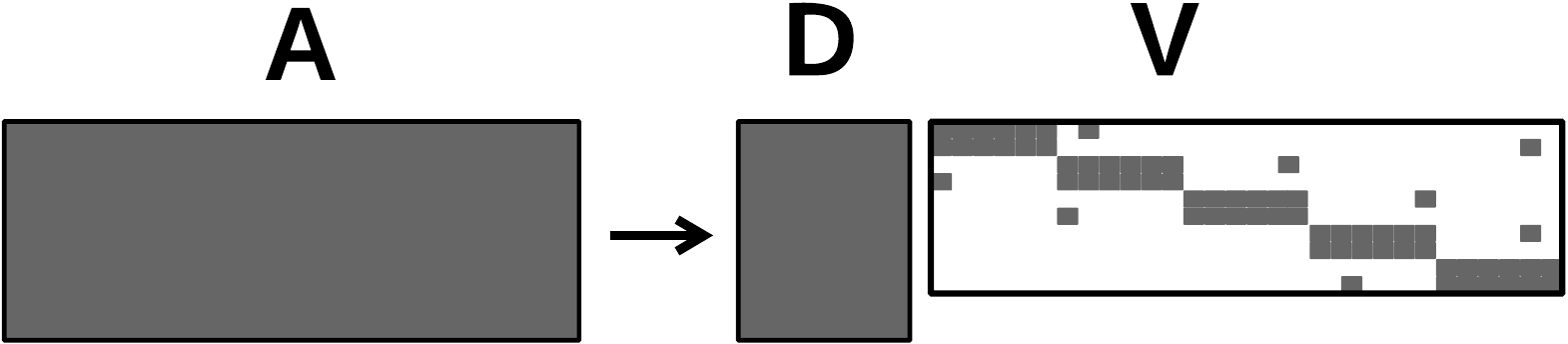}
	\caption{ {\em Schematic of decomposing a dense data matrix into the product of a small dense matrix and a large sparse matrix.}} \label{fig:densesparse1}
\end{figure}

When $\G$ is massive and dense, each distributed update in (\ref{eq:updategeneral}) becomes very expensive. To cope with large data sizes, \sys{} creates an approximation to $\G$, denoted by $\widehat{\G}$, to reduce the cost of an update. To be more specific, our aim is to decompose the data matrix $\A$ into two components, i.e., $\widehat{\A} = \D\V$, where $\D \in \R^{m\times l}$ contains a subset of columns from $\A$, $\V \in \R^{l \times n}$ is a sparse matrix, and ${\rm rank}(\A) \le l \ll n$ (see Figure \ref{fig:densesparse1}). After decomposing $\A$, we then efficiently partition the decomposed data and perform distributed updates using $\widehat{\G}=(\D\V)^T(\D\V)$. Mapping the original dense data to a decomposed model directly reduces the memory usage, and the costly computational operations and communication incurred by the iterative updates. 

When $\A$ is low-rank, it is possible to construct a reduced decomposition that is exact ($\A = \D\V$). However, we demonstrate that for many real-world datasets, we can achieve significant performance improvements in exchange for a small decomposition error ($\A \approx \D\V$). We discuss the connection between the decomposition error and the accuracy of a target learning method as well as strategies for tuning the decomposition to achieve a desired level of accuracy in the iterative learning algorithms. 

\begin{figure*}
	\centering
	\includegraphics[width=.88\textwidth]{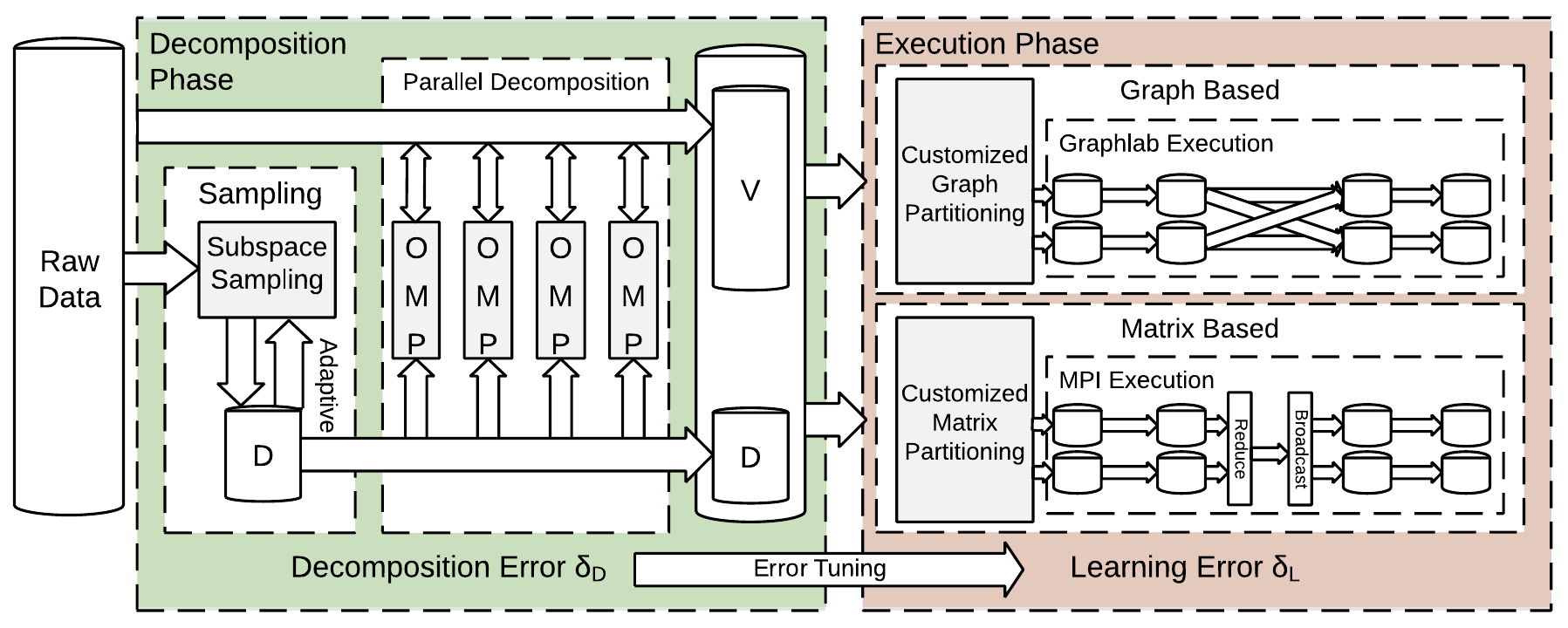}
	\caption{{\em An overview of \sys{} framework.} The method is divided into two main phases, the decomposition phase (left) and the execution phase (right). To execute iterative updates, we provide two computational models: a matrix-based model (implemented in MPI) and a graph-based model (implemented in GraphLab).}\label{fig:sys}
\end{figure*}

\sys{} consists of three main components (see Figure \ref{fig:sys}): (i) A scalable data decomposition that shrinks the size of the data set by leveraging the data's structure, (ii) A data partitioning scheme along with an execution flow for performing iterative updates on the decomposed data that significantly reduces the distributed computing costs, (iii) A systematic method for tuning the decomposition error (denoted by $\delta_D$) to achieve the desired level of approximation error in the learning algorithms (denoted by $\delta_L$).

\subsection{Target applications} \label{sec:targetapps}
Our framework can be used for a broad class of optimization problems that are solved via iterative updates based upon the Gram matrix. A large number of objective functions used in machine learning, e.g., penalized regression methods such as the LASSO or BPDN \cite{chen1998atomic}, and ridge regression \cite{hoerl1970ridge}, are typically solved using iterative updates. In all these settings, the complexity of executing these methods is dominated by costly iterative computation on the Gram matrix of the dataset.

To ground \sys{} in real-world problems, we now discuss two particular learning algorithms that are evaluated in this paper: (i) sparse approximation and (ii) the power method for eigenvalue decomposition.\\

\noindent \textbf{(i) Sparse approximation for image denoising and classification.} Sparse representation is used in a wide range of signal processing and machine learning applications, including denoising \cite{chen1998atomic}, classification \cite{wright2010sparse}, clustering \cite{DyerJMLR13}, and outlier detection \cite{jour:dyerarxiv15}. The sparse approximation objective function can be written in terms of the $\ell_1$-norm as follows: 
\begin{equation}
\label{eq:l1}
\underset{\x}{\arg\min}~ \|\A\x-\y\|_2 + \lambda \|\x\|_1, 
\end{equation}
where $\x$ is a sparse representation of $\y$ with respect to $\A$ and $\lambda$ is a regularization coefficient (increasing this parameter promotes sparser solutions). 

In an image denoising application, $\y$ is a noisy image, $\x$ is a sparse coefficient vector, and $\A\x $ is a denoised approximation of $\y$. In a classification application, the sparse coefficient vector $\x$ is used to determine which class a test signal $\y$ belongs to. This can be done by first measuring the sum of the coefficients in each class and then finding the class that has largest number of nonzero coefficient.

In the sequel, we evaluate the performance of RankMap for accelerating first-order methods for sparse approximation via $\ell_1$-minimization. This sparse approximation problem can be solved using the following iterative soft thresholding (IST) algorithm \cite{daubechies2004iterative}:
\begin{eqnarray}\label{eq:update}
\x^{iter+1}= f(\x^{iter} - \gamma ( \G \x^{iter}- \A^T\y)),
\end{eqnarray}
where $f(.)$ is a low-complexity thresholding operation (e.g., a soft-thresholding operator \cite{daubechies2004iterative}) to account for the term $\lambda \|\x\|_1$ at each iteration, and $\gamma$ is the step size. In our evaluations, we employ a variant of this algorithm called FISTA \cite{jour:Beck09}. FISTA is an example of a projected gradient descent (PGD) methods \cite{Lin:2007:pgd} which provide a generalization of standard gradient descent methods for certain classes of non-smooth objective functions. \sys{} can be readily applied to cost functions that can be solved using PGD.\\

\noindent \textbf{(ii) Power method for eigenvalue decomposition.} The power method is a simple and iterative algorithm that can be used to sequentially find the eigenvectors and eigenvalues of a matrix in descending order. Recall that an eigenvector $\x$ of a matrix $\A$ satisfies the following relationship $\A\x=\sigma \x$, where $\sigma$ is the eigenvalue  associated with the eigenvector $\x$. To find an eigenvector of the symmetric matrix $\G = \A^T \A$, the power method utilizes the following iterative update: 
\begin{eqnarray}\label{eq:updatepower}
\x^{iter+1}=\frac{\G\x^{iter}}{\|\G\x^{iter}\|_2}.
\end{eqnarray}
Once the power method converges to an estimate of an eigenvector $\x$, the contribution of this eigenvector is removed from $\A$, and the power method is applied again to the residual to find the next eigenvector. 

In both applications described above, the main cost of each iteration is due to the computation of $\G \x$, especially when $\G$ is large, dense, and distributed onto multiple computing nodes. For example, as a case-study in our evaluations, we perform image reconstruction on a dataset where $\A$ is a collection of light field image patches of size $18,496\times100,000$. In this case, to reconstruct a single noisy image patch $\y$, more than $3.6$ billion floating point multiplications are required to perform $\G\x^{iter}=\A^T\A\x^{iter}$ per iteration.

\section{Background and Related Work}\label{sec:related}
In this section, we provide background on methods for matrix factorization and describe related work.

\subsection{Methods for matrix factorization }\label{sec:reflow}
High-dimensional data can be modeled by the low rank structures that are present in the data. Extracting low dimensional structures not only reduces dimensionality, but also mitigates the effect of noise and improves the performance of learning and inference tasks \cite{jour:dyerarxiv15,vidaljournal}.

\subsubsection{Singular value decomposition (SVD)}
In settings where the column span of $\A$ admits a low rank model, the SVD provides a powerful tool for forming low rank approximations. Let $\A = \U {\bf S} \V^T$ be the SVD. The best rank-$k$ approximation of $\A$ is given by $\A_k = \U_k {\boldsymbol \Sigma}_k \V_k^T,$ where $\U_k \in \R^{m \times k}$ and $\V_k \in \R^{n \times k}$ are the truncated left and right singular vectors (first $k$ columns of $\U$ and $\V$) and ${\boldsymbol \Sigma}_k \in \R^{k \times k}$ contains the first $k$ singular values of $\A$ along its diagonal. The rank of $\A_k$ equals the number of non-zero singular values. The truncated SVD also provides the solution to principal components analysis (PCA), which seeks to find a $k$-dimensional subspace that best approximates $\A$ in the least-squares sense \cite{thompson72}. 

The complexity of computing the SVD directly is $m^2n$. Thus, for large datasets, the power method is used to find the eigenvectors of $\A^T \A$ and $\A \A^T$, which correspond to the right and left singular vectors respectively. 

\subsubsection{Sparse factorization}
The SVD provides a closed-form solution for finding the best rank-k approximation to a matrix. However, in many settings, enforcing sparse structure, either in the left or right singular vectors can provide a more faithful and compact decomposition of the data. Two widely used sparse factorization methods include sparse PCA (SPCA) \cite{jour:zou2006} and dictionary learning (DL) \cite{ksvd}. However, these approaches are often not applied to large datasets since computing an update of both the left and right factor matrices, at each iteration is costly. To solve SPCA on big datasets, a generalized power method can be employed \cite{journee2010generalized}. The basic idea behind using the power method to find sparse principal components is to simply threshold the output of each power iteration to ensure the resulting eigenvectors are sparse. Unfortunately, the convergence of this method is much slower than standard power iterations.

\subsubsection{Column subset selection (CSS)-based matrix factorization} 
An alternative strategy for low rank matrix factorization is to form a decomposition based upon columns (or rows) from the data. CSS-based solutions form an approximate matrix decomposition in which one factorized component is a subset of the columns of the data itself \cite{drineas2005nystrom}. CSS-based approaches have been used to provide a scalable and efficient strategy for finding approximate solutions to least-squares regression \cite{jour:Drin04}, Gramian matrix decomposition \cite{drineas2005nystrom}, image denoising and clustering \cite{jour:dyerarxiv15}, and also in spectral clustering \cite{fowlkes2004spectral}. After selecting columns from $\A$, the remaining unsampled columns are completed by finding the least-squares projection onto the subspace spanned by the sampled columns. 

\subsection{Generic distributed abstractions } 
A number of successful distributed abstractions for processing large datasets on clusters have been proposed. Examples include MapReduce \cite{jour:dean2008}, Apache Spark \cite{zaharia2010spark}, and SystemML \cite{systemml11}. However, these models become less efficient for applications when direct data-parallelism does not exist. Several new distributed abstractions have been proposed that model data dependency in a graph format, most notably
Pregel \cite{jour:malewicz2010} and GraphLab \cite{jour:low2010}. They use a vertex-centric
computation model, in which the user-defined programs are executed on each vertex in parallel. 
As graph-based abstractions are suited for sparse datasets, efficient data partitioning is not possible when the graph-representation of the data is densely connected. Furthermore, such tools mostly rely on the communication between the vertices for computation. When the data is densely connected, the resulting communication congestion makes the computation dramatically slow. Because of this, most of these tools are designed based on the assumption that the input data is sparse \cite{OSDI:Gon12, jour:malewicz2010, zaharia2010spark}.

By design, MapReduce-based solutions are not guaranteed to be fast, instead they provide easy and reusable programming frameworks that operate on very large datasets on a distributed computing platform. Users only have to deal with writing the functions of the algorithm in the given MapReduce-based programming model. MapReduce, on the other hand, controls the distributed cluster, manages data partitioning and data transfers between the various parts of the system, and provides fault tolerance.

\section{Column selection-based sparse decomposition (CSSD) }\label{sec:factor}
In this section, we present a scalable method for matrix decomposition (the Decomposition phase in Figure \ref{fig:sys}) which we call Column Selection-Based Sparse Decomposition (CSSD).

\subsection{Overview of CSSD method}
The main idea behind CSSD is to first select a subset of columns from $\A$, and then use this subset of columns as a basis from which we form sparse representations of the remaining columns. We thus factorize the data as $\A = \D \V$, where $\D$ is formed by subsampling and normalizing the selected columns of $\A$. Each column of $\V$ is then computed by finding the sparse approximation of the corresponding column of $\A$ with respect to $\D$. This sparse approximation problem can be solved by an efficient greedy routine called \textit{orthogonal matching pursuit} (OMP) \cite{davisOMP}. We provide pseudocode for CSSD in Algorithm \ref{alg:smac}.

\subsubsection{Step 1. Sequential column selection}
In order to ensure that the total approximation error in our factorization is sufficiently small, we must ensure that the columns selected from $\A$ to form $\D$ well approximate the range of the original matrix. Thus, we employ a sequential method to adaptively select columns that are not well approximated by the current set of columns \cite{deshpande2006matrix}.

Adaptive column selection methods select a new batch of columns according to the following probability distribution:
\begin{equation}
\label{eq:pdf}
p(i) \propto \frac{\|\A_S \A_S^+ \a_i - \a_i\|_2}{\|\a_i\|_2},
\end{equation}
where $p(i)$ equals the probability of selecting the $i^{\rm th}$ column from $\A$ (denoted by $\a_i$), $S$ contains the indices of columns already selected, and $\A_S$ denotes the sub-matrix of sampled columns. We can flexibly execute this subsampling approach by either specifying the maximum number of columns to select and/or specifying the maximum amount of error in each unsampled column of $\A$.

\begin{algorithm}[t!]
 \caption{: {\bf Column Selection Sparse Matrix Decomposition}}
 \label{alg:smac}
\begin{algorithmic}
 \STATE {\bfseries Input:} Matrix $\A \in \R^{m \times n}$, error tolerance $\delta_D$, number of columns to select at each iteration $l_s$, and the maximum number of columns to select $l$.
 \vspace{1mm}
 \STATE {\bfseries Output:} A sparse matrix $\V \in \R^{l \times n}$ and a dense matrix $\D\in \R^{m \times l}$ such that for each column of $\a$ of $\A$, $\|\a-\D\v\|_2\le\delta_D$.
 \vspace{1mm}
 \STATE {{\bfseries Initialize:} Initialize $\D$ by adding and normalizing $l_s$ columns from $\A$ with uniform random sampling.}
 \vspace{2mm}
 \STATE{{\bfseries \underline{Step 1: Sequential column selection}}}
 \vspace{1mm}
 \WHILE{$ncols( \D) < l$}
 \STATE{ I.~ Update $\D$ by selecting and normalizing $l_s$ columns \\\hspace{3mm} from $\A$ according to the distribution in (\ref{eq:pdf}).}
 \STATE{II. If the $\ell_2$-norm of each column of $\E = \A - \D \D^{+} \A$ \\\hspace{3mm} is less than $\delta_D$, return $\D$ and proceed to Step 2 to \\\hspace{3mm} compute $\V$.}
 \ENDWHILE
 \vspace{2mm}
 \STATE{\bfseries{\underline{Step 2. Sparse approximation}}}
 \vspace{1mm}
\STATE{ I. Compute $\V$ by applying Batch OMP to solve (\ref{eq:adv})\\\hspace{3mm} with error tolerance $\delta_D$.}
\end{algorithmic}
\end{algorithm}

\subsubsection{Step 2. Sparse approximation}
After selecting a subset of $l$ columns $\A_S \in \R^{m \times l}$, we normalize each column such that all the columns in matrix $\D$ have unit norm. Now, to form the sparse matrix $\V$, we find a sparse representation of the remaining columns in $\A$ (i.e., $\A_{-S}$) in terms of the normalized dictionary $\D$. The problem is formally written as follows:
\begin{equation}\label{eq:adv}
\quad \min_{\v} \quad \| \v \|_0 \quad \text{s.t.} \quad \frac{\| \a_i - \D \v \|_2}{\| \a_i \|_2} \le \delta_D, ~\forall i \notin S.
\end{equation}

\noindent where $\| \v \|_0$ counts the number of nonzero coefficients in $\v$ and $\delta_D$ is a user-specified parameter which controls the decomposition error.

 We employ a matching pursuit-based solver called Batch OMP \cite{rubinstein2008efficient} to solve (\ref{eq:adv}). We can enforce sparsity either by the number of non-zeros in each column of $\V$ (i.e., $\|\v\|_0$) or by the total amount of approximation error for each column.

\subsection{Complexity analysis}
The complexity of sequential column selection (Step 1) is $\mathcal{O}(l^2m+lmn)$. The complexity terms correspond to computing $\D^+$ and $\D\D^+\A$ respectively. The projection $\D\D^+\a$ can be computed for each column of $\A$ independently. The complexity of sparse approximation (Step 2), using the Batch OMP method \cite{rubinstein2008efficient}, is $\mathcal{O}(lmn+{k}^2ln)$, where $k<l$ is the average number of non-zeros per column of $\V$. Similarly, for each column of $\A$, Batch OMP is applied independently. Let $n_c$ be the number of parallel processing nodes. By storing $\D$ (which is a small $m\times l$ matrix) and a uniform fraction of columns of $\A$ in each node (i.e., $\frac{n}{n_c}$ columns), the overall complexity of Algorithm 1 in a distributed setting can be written as $\mathcal{O}(\frac{n}{n_c}(lm+{k}^2l)+l^2m)$.

Note that CSSD is linear in terms of both the number of data samples $n$, and the number of processors $n_c$. This is a key feature of our approach that makes our framework applicable to very large datasets in distributed settings.

\subsection{Computational benefits of CSSD}
CSSD provides computational benefits when the size of the decomposition is small (i.e., $l$ is small relative to $m$) and/or when matrix $\V$ is sparse. In general, predicting
the amount of savings in computation is a function of (i) the structure of the data and (ii) the amount of accuracy required from the learning algorithm. We now discuss some key factors that impact the decomposition results.\\

\vspace{-1mm}
\noindent {\bf Impact of data structure.}
Predicting the size and sparsity of the decomposition provided by CSSD for an arbitrary dataset is challenging; however, when the data lies on a single subspace (i.e., exhibits low rank structure) or lies on multiple low-dimensional subspaces, CSSD provides a more compact representation of the data. For example, when data is exactly low rank and its rank is $r<m$, we must select $r$ linearly independent columns from $\A$ to form an exact decomposition (zero error), i.e., $l=r$. When the data is approximately low rank, there exists a large body of work that characterizes the performance of the sequential column selection method (Step 1) used to form $\D$ \cite{deshpande2006matrix,ICML:Git13}. In particular, the selection strategy in Step 2 of Algorithm \ref{alg:smac} provides exponential decrease in the factorization error with each batch of columns that we select from $\A$ \cite{deshpande2006matrix}. More specifically, assume that at each iteration, we select $l_{\rm s} > \frac{r}{\epsilon}$ samples from the columns of $\A$ and let $l = t l_{s}$ denote the set of columns selected after $t$ iterations. Let $\A_r$ denote the best rank $r$ approximation to $\A$ and let $\widetilde{\A} = \A_S\A_S^{+} \A$ denote the approximation of $\A$ based upon the $l$ selected columns $\A_s$. Then according to \cite{deshpande2006matrix}, the difference between the expected value of the approximation error, i.e., $\| \A - \widetilde{\A} \|_F^2$ and that of the best rank $r$ approximation $\|\A - \A_r \|_F^2$ decreases exponentially with rate $\epsilon^t$.

Another low-dimensional signal model that has recently gained traction models data with multiple low-dimensional subspaces (union of subspaces). For example, images of objects under different illumination conditions \cite{facesubs}, motion trajectories of point-correspondences \cite{kanatani01}, neural data \cite{jour:dyerarxiv15}, to structured sparse and block-sparse signals \cite{modelcs} are all well-approximated by a union of low-dimensional subspaces. When $\A$ lies on a union of subspaces, this effectively bounds the sparsity level of each column of $\V$ \cite{DyerJMLR13}. This insight is based upon the fact that when we form a representation of a column of $\A$ with respect to other columns in the same dataset (as in CSSD), the sparsity level of each column is bounded by the dimension of the subspace the signal lies on. For instance, if $\A$ lives on a union of multiple $r$-dimensional subspaces of $\R^n$ and we select at least $r$ linearly independent columns from each subspace, then no more than $r$ non-zeros are required to represent a signal. In other words, the number of non-zeros per column of $\V$ is no more the dimension of the subspace the signal lives in. \\

\vspace{-1mm}
\noindent {\bf Impact of increasing the decomposition error.}
\label{sec:compact}
The decomposition error of CSSD is controlled by the parameter $\delta_D$ in Algorithm \ref{alg:smac}. In the case where we set $\delta_D = 0$, then we are guaranteed an exact decomposition of the data. Exact decomposition occurs when $r$ linearly independent columns are selected from $\A$, where $r = \rm{rank}(\A)$. In this case, the selected columns in $\A_S$ will fully represent the data and thus $\| \A - \A_S \A_S^{+} \A \| = 0$, i.e., exact decomposition occurs.

While CSSD can produce an exact decomposition when the data is exactly low rank (or lies on a union of subspaces), in practice, datasets are approximately low rank. In this case, we can introduce a small amount of error into the decomposition by setting $\delta_D > 0$. By introducing some error into the decomposition, we observe that both the number of selected columns in Step 1 of Algorithm \ref{alg:smac} and the sparsity level of $\V$ can be reduced further. In Figure \ref{fig:powermethod}, we show how increasing the decomposition error produces a more compact decomposition. In Section \ref{sec:impact}, we study and discuss the impact of the decomposition error $\delta_D$ on the accuracy of a learning algorithms that we apply \sys{} to.

\section{Tuning decomposition error for a target learning accuracy}\label{sec:impact}

In the previous section, we discussed the computational benefits associated with introducing some approximation error into a CSSD decomposition. Naturally, as we increase the decomposition error (controlled by $\delta_D$), the accuracy of our learning algorithm will be affected. Thus, the key question is how much decomposition error we can afford to achieve a certain degree of accuracy in learning. The answer to this question heavily depends on the specific learning algorithm and the application of interest.

Previous theoretical studies have established a connection between the total error in a factorization of a kernel (or Gram) matrix and the accuracy of certain popular learning algorithms, including: kernel ridge regression and kernel SVM \cite{cortes2010impact}. While for some learning algorithms, our framework can exploit previous work to relate $\delta_D$ and the learning error which we denote by $\delta_L$, the aim of this section is to propose a generic approach for tuning the factorization error to achieve a specified learning accuracy. We do this by iteratively remapping of the data to find a compact decomposition that satisfies a learning error (specified by the user).

\subsection{Error tuning}
Given an already established relationship between the decomposition error and a specific algorithm, a practitioner who uses our framework can easily specify the error parameter $\delta_D$ for CSSD to achieve a particular learning accuracy. Alternatively, if a practitioner specifies a target accuracy for a learning algorithm, the decomposition error $\delta_D$ can be tuned in order to achieve a particular learning error $\delta_L$.

Our strategy for guaranteeing that we have small learning error, is to solve CSSD for a particular $\delta_D$, map the resulting decomposition via the methods described in Section \ref{sec:fact}, and then compute the accuracy of the learning algorithm $\delta_L$. We then iteratively add columns to $\D$ such that the decomposition error $\delta_D$ is small enough to ensure that $\delta_L$ is within the error-specified tolerance. Depending on the underlying computing resources available, \sys{} can be applied for multiple values of $\delta_D$ in parallel and the largest value of $\delta_D$ (most compact representation) that achieves a particular value of $\delta_L$ is selected.

When computing resources are constrained and thus running the algorithm for multiple values of $\delta_D$ in parallel is not possible, we use a bisection method. In essence, the idea is to: (i) set the factorization error to predefined maximum value $\delta_D^{max}$ (0.4 in our experiments) and evaluate $\delta_L$, (ii) if $\delta_L$ is below a target value then we stop, otherwise we decrease $\delta_D$ by half. By exponentially decreasing $\delta_D$, we are also guaranteed to decrease $\delta_L$ exponentially, provided that there is a polynomial relationship between the two quantities. We observe a polynomial relationship holds both in theory \cite{cortes2010impact} and in practice. In Section \ref{sec:eval}, we provide empirical results which demonstrate the connection between the decomposition error and learning accuracy for numerous datasets and algorithms of interest (see Figures \ref{fig:face_c} and \ref{fig:powerdeltal}).

\section{Distributed execution and data partitioning}\label{sec:fact}
In this section, we introduce our approach for applying
iterative updates on the decomposed data (the execution phase in Figure \ref{fig:sys}). We describe an execution flow for dependency-matrix based updates (i.e., $\G\x=\V^T (\D^T\D) \V\x$) and introduce an efficient method for partitioning the decomposed data in a distributed setting. We also provide performance bounds on memory usage, the number of flop operations, and the number of communicated bytes across the computing nodes.

\subsection{Computation flow}
We propose two computational models for the distributed implementation of an update in (\ref{eq:updategeneral}). Recall that at each iteration, we must compute $\z = \G\x = \V^T(\D^T \D)\V\x$. We break this computation into four steps\TODO{eva: use different variable for $p$ to avoid conflict}: (i) $ \p= \V\x $ (ii) ${\bf r} = \D\p$, (iii) $ \p= \D^T{\r}$, and (iv) $ \z = \V^T\p$. The output vector $\z$ is used to produce an update of $\x^{{\rm iter}+1} = f( \z + \b),$ where $\b$ is an offset vector, and $f(\cdot)$ is a low-complexity function such as a soft-thresholding operator (sparse approximation) or normalization (power method). To carry out the computation described above, we propose and implement a matrix-based and vertex-based model to apply the iterative updates on the decomposed factors. We now describe our implementation of both models.

\subsection{Matrix-based model}
\begin{figure}[t!]
	\centering
	\includegraphics[width=.34\textwidth]{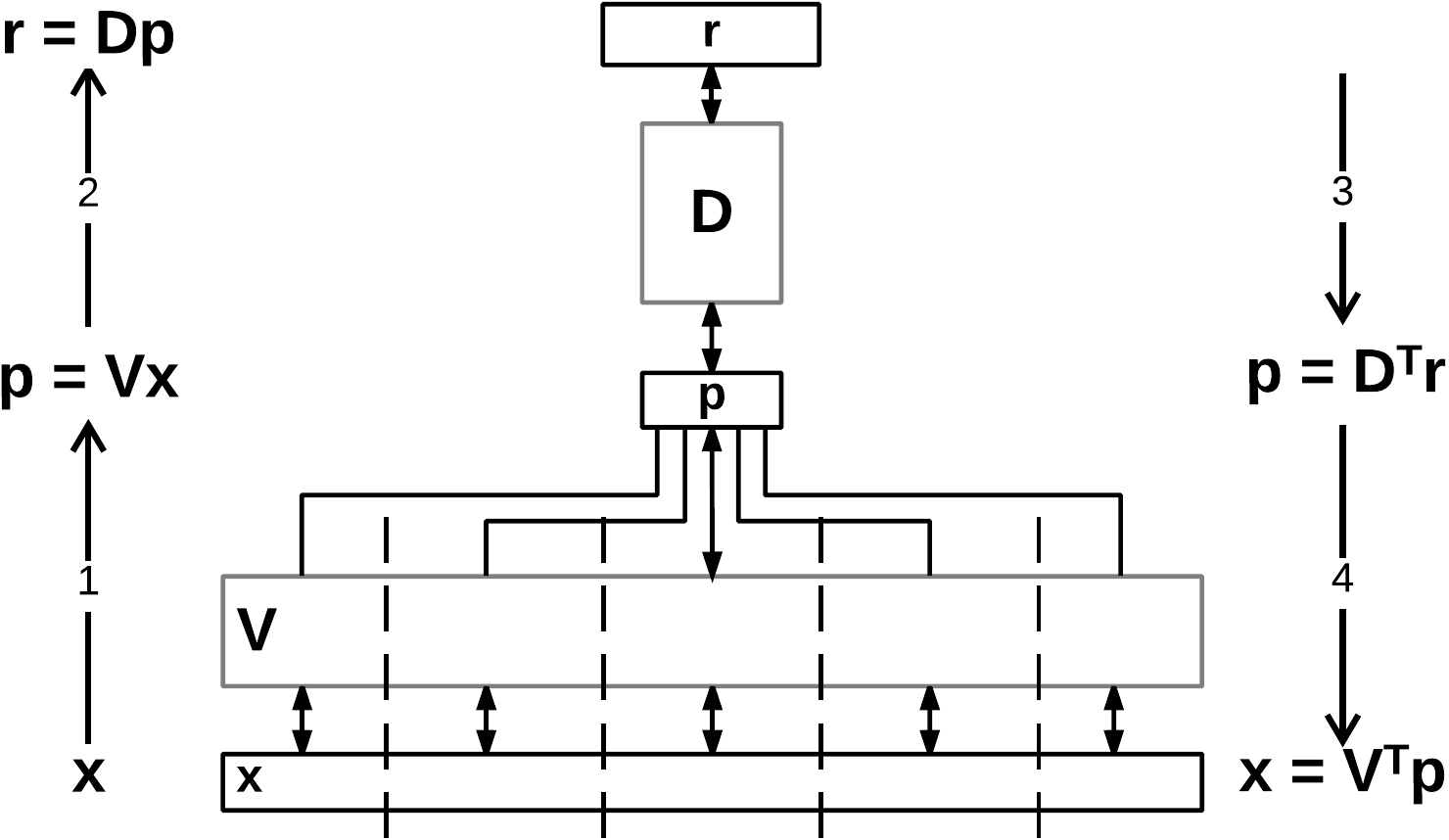}
	\caption{{\em Distributed design of matrix-based model.}}\label{fig:densesolver}
\end{figure}

Figure \ref{fig:densesolver} shows the schematic of our proposed matrix-based model. In this model, the data is stored in arrays. The sparse matrix $\V$ is stored and operated upon using the Compressed Sparse Column (CSC) format. The matrix $\D$ and vector $\x$ are stored using regular dense arrays. By doing so, we exploit sparsity in $\V$. We use C++ Eigen Library for array manipulation and MPI for distributed computing.

\subsubsection{Distributed partitioning}
We partition columns of $\V$ uniformly across the computing nodes to achieve a balanced partitioning. Let us assume that there are $n_c$ computing nodes. Thus, $\frac{n}{n_c}$ number of columns are assigned to each node. The vector $\x$ is also divided into chunks of size $\frac{n}{n_c}\times1$. Each chunk is then allocated to the node that hosts the corresponding columns of $\V$.

Matrix-vector multiplications $\V\x$ are performed locally on the columns of $\V$ and the portion of $\x$ that reside on the same computing node. The resulting $l\times1$ vectors are then sent to a central node to create $\p=\V\x$. Next, $\D^T(\D\p)$ is computed locally in the central node. The resulting $l\times1$ vector is then \textit{broadcasted} back to all the computing nodes where it is multiplied by the local $\V^T$ to update the vector $\x$.

\subsubsection{Performance bounds}
We now provide bounds on the memory usage, computation, and communication required by our proposed matrix-based model. We also provide the performance bounds for baseline, in which we perform iterative analysis on $\A^T\A$ in matrix format. Let \textit{nnz(.)} denote the number of non-zeros of its input and $n_c$ denote the number of computing nodes. Recall that $\D$ is a $m \times l$ matrix and $\V$ is a $l \times n$ matrix. Note that in our target data scenarios $m\ll n$ and $l\ll n$. In many cases, the rank of decomposition $l$ is often much smaller than the dimensions of data $m$. When data exhibits union of subspace property, $\V$ will be sparse, i.e., $nnz(\V)<ln< mn$. 

\begin{center}\scalebox{.9}{
\begin{tabular}{|c|c|c|}
\hline
 Matrix-based & Baseline & RankMap    \\ \hline
Memory usage $\propto$:  & &
\\ 
\# non-zeros  & $mn$ & $(nnz(\V)+ lm)+n+m$
\\ \hline
Computation $\propto$ : & &\\
\# additions  & $2mn+ m n_c$ &  $2(nnz(\V)+ l m+l n_c)$ \\
\# multiplications  & $2mn$ &$2 (nnz(\V)+ l m)$       
\\ \hline
Communication $\propto$: & &\\
 \#edges  & $2lm$ & $2 l n_c$        
 \\ \hline  
\end{tabular}}
\end{center}

Since $\V$ is stored in a CSC format, only the non-zero values are stored and operated on. The matrix $\D$ is stored in a regular dense matrix format. The communication corresponds to sending and receiving the $l\times1$ vectors from each computing node to the central node.Clearly, for smaller $l$ and sparser $\V$, both memory footprint and the number of arithmetic operations are reduced. The number of edges, which correspond to the number of broadcasted and reduced values, directly corresponds to $l$ and the number of computing nodes $n_c$.

\subsection{Graph-based model}
\label{sec:graphmodel}
\begin{figure}[t!]
	\centering
	\includegraphics[width=0.44\textwidth]{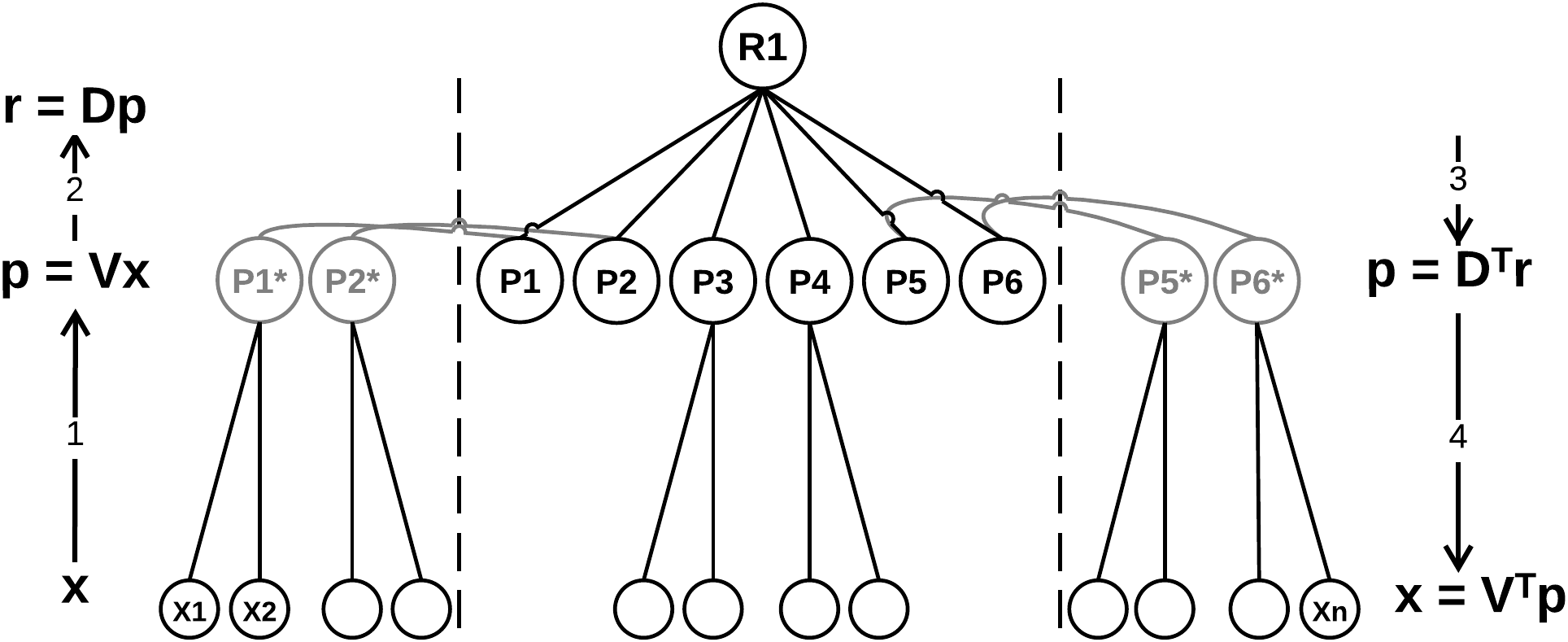}
	\caption{{\em Distributed design of graph-based model.}}\label{fig:sparsesolver}
\end{figure}

Figure \ref{fig:sparsesolver} shows an schematic of our proposed graph model. The decomposed data is three-layer graph denoted by $\mathcal{G_A}(S_X,S_P,S_R)$ with vertex sets $S_X=\{X_i\}_{i=1}^{n}$ in the bottom layer, $S_P=\{P_i\}_{i=1}^l$ in the middle layer, and $S_R=\{R_1\}$ in the top layer. Each non-zero element in $\V$, e.g., $\V_{ij}$, is represented by an edge which connects $X_i$ to $P_j$. Each column of $\D$, e.g., $\D_{i}$, is represented by an edge which connects $P_i$ to $R_1$. Value of vertices in $S_X$ correspond to the elements of $\x$.

We use GraphLab Distributed API \cite{jour:low2010} to implement this model. While GraphLab is a highly optimized distributed engine for Graph-based computation on iterative data, we perform extensive customizations in order to adapt GraphLab to our factorized setting. We also force GraphLab to use our developed graph partitioning method as opposed to its automated partitioning schemes. Our proposed partitioning is customized to the factorized data and significantly improves the performance.

\subsubsection{Distributed partitioning}
In the graph-based model, we partition $\mathcal{G_A}(S_X,S_P,S_R)$ with the aim of balancing the number of components assigned
to each node and also minimizing the inter-node
communications characterized by the edges. Since the edge distribution of $\mathcal{G_A}$ is highly non-uniform ($l\ll n$), a vertex partitioning inevitably results in many undesirable edge-cuts across the computing nodes. Instead, we apply a vertex-cut method in which the goal is to partition graph edges evenly such that the number of vertices that are spanned across multiple partitions is minimized. As a result of edge partitioning, vertices may reside on two or more computing nodes. In this case, we assign one of the copies to be the \textit{master} vertex and the others to be the \textit{replica} vertices (these definitions are borrowed from GraphLab \cite{OSDI:Gon12}). The replicas directly cause (expensive) inter-node communication costs.

Figure \ref{fig:sparsesolver} shows the graph-based distributed design. Our detailed edge partitioning method is as follows. (i) Distribute master of vertices $X_i\in S_X$ uniformly onto the available computing nodes such that vertex chunks of size $\frac{n}{n_c}$ are assigned to each node. (ii) Add the edges between vertices $X_i\in S_X$ and $P_j\in S_P$ to the node in which the corresponding master of $X_i$ resides. (iii) Add master of vertices $P_i \in S_P$ and $R_1\in S_R$ to a \textit{central node}. (iv) Add the edges between the vertices $P_i\in S_P$ and $R_1\in S_R$ to the central node.

The proposed edge partitioning algorithm is highly efficient in that it does not induce any replicas for vertices in $S_X$ and $S_R$. However, from Step (ii), replicas of vertices in $S_P$ may exist in computing nodes other than the central node. At the beginning of an iteration, master vertices in $S_P$ and their replicas perform vertex updates with respect to $S_X$. The replicas send the updated values to their own master vertices in the central node. The master vertices in $S_P$ \textit{reduce} the received values ($\p=\V\x$). Then master vertex $R_1$ performs a vertex update ($\r=\D\p-\y$). Next master vertices in $S_P$ complete vertex updates with respect to $S_R$ and \textit{broadcast} the results to their own replicas ($\p=\D^T\r$). Finally, master vertices in $S_X$ update themselves ($\x=\V^T\p$). We integrate and implement the proposed customized partitioning
and distributed computation flow with the distributed GraphLab API \cite{OSDI:Gon12}.

\subsubsection{Performance bounds}\label{ssec:graphper}
We now provide bounds on the memory usage, computation, and communication required by our proposed graph-based model. 

\begin{itemize}[noitemsep]
\item {\bf Memory usage} \hfill \\
\# edges $\propto\textit{nnz}(\V)+l$.\hfill \\
\# vertices $\propto n+\sum_{1\leq i\leq l} rep(P_i)$.
\item {\bf Computation} (per iteration) \hfill \\
\# additions $\propto 2(\textit{nnz}(\V)+ml)+\sum_{1\leq i\leq l}rep(P_i)$. \hfill \\
\# multiplications $\propto2(\textit{nnz}(\V)+ml)$.
\item {\bf Communication} \hfill \\
\# edge-cuts $\propto 2\sum_{1\leq i\leq l} rep(P_i)$.
\end{itemize}

\begin{center}\scalebox{.8}{
\begin{tabular}{|c|c|c|}
\hline
 Graph-based & Baseline & RankMap    \\ \hline
Memory usage $\propto$:  & &
\\ 
\# edges  & $mn$ & ${nnz}(\V)+l$\\
\# vertices  & $n+mn_c$& $n+\sum_{1\leq i\leq l} rep(P_i)$
\\ \hline
Computation $\propto$ : & &\\
\# additions  & $2mn+ m n_c$ &   $2(\textit{nnz}(\V)+ml)+\sum_{1\leq i\leq l}rep(P_i)$ \\
\# multiplications  & $2mn$ &$2 (nnz(\V)+ l m)$       
\\ \hline
Communication $\propto$: & &\\
 \#cuts & $2lm$ & $2\sum_{1\leq i\leq l} rep(P_i)$        
 \\ \hline  
\end{tabular}}
\end{center}

Each of the computing nodes receive approximately $\frac{1}{n_c}(n+\sum_{1\leq i\leq l}~rep(P_i))$ vertices and $\frac{1}{n_c}nnz(\V)$ edges. The central node has $l$ additional edges between the master vertices in $S_P$ and $R_1$. The computation cost is induced by vertex update operations. The communication overhead is incurred by the message passing across master and replica vertices in $S_P$.

\textbf{Bound on total replicas.} From the discussions above, it is clear that reducing number of replicas of $S_P$ reduces the communication overhead. The following are the bounds on the total number of replicas:

$$l\leq \sum_{1\leq i\leq l}~rep(P_i)\leq l n_c.$$ 

The inequalities hold since each $P_i$ is replicated at least once and at most $n_c$ times (one replica per computing node). Both $l$ and $n_c$ are much smaller than the size of the graph. Thus, \sys{}'s graph-based model readily provides efficient/balanced computation and reduced communication without using complicated and costly graph partitioning algorithms. The minimum communication is achieved when $\V$ is block-diagonal.

\section{Evaluations}\label{sec:eval}
In this section, we evaluate the performance of \sys{} on a variety of datasets. Our evaluations explore: (i) the scalability of CSSD and its ability to produce sparse representations, (ii) the connection between decomposition error and learning accuracy for multiple learning applications including face recognition, image denoising, and PCA, (iii) \sys{}'s performance improvement in terms of runtime, and memory over prior work, and (iv) the performance of our distributed matrix- and graph-based models for different structured data sets.

\subsection{Evaluation setup}

\subsubsection{Datasets}
We evaluate \sys{} on both real and structured synthetic datasets. Our real datasets include Light Field data \cite{lfweb}, hyper spectral images \cite{salina}, a dictionary of video frames \cite{hitomi2011video}, and a collection of images of different faces under varying illumination conditions \cite{yaleb}.

We apply \sys{} to two different Light Field datasets. The first dataset, which we refer to as Light Field (i), consists of $10k$ randomly selected atoms from a $5\times5$ Light Field array (collected from Chess Images). Each Light Field patch consists of 25 $8\times8$-patches which produces a dataset of size $1.6k\times10k$ ($128MB$). The second dataset, which we refer to as Light Field (ii), consists of $100k$ randomly selected atoms from a $17\times17$ Light Field array (collected from all available Light Fields in the archive). Each Light Field patch consists of 289 $8\times 8$-patches which produces a dataset of size $18496\times100k$ ($14.7GB$). The hyper spectral dataset (Salinas) is taken from a region of a remote sensing scene in Salinas, CA. Each pixel in the scene contains information from $203$ spectral bands and produces a dataset of size $203\times54129$ ($87.9MB$). The video dictionary dataset (VideoDict) contains patches of an image over multiple frames and produces a dataset of size $1764\times100000$ ($ 1.41GB$). The face image dataset (Faces) consists of $631$ images of 10 different peoples faces under varying illumination conditions. Each image is $48 \times 84$ pixels, which produces a dataset of size $4032 \times 631$. In addition to real-world datasets, we generate synthetic data for $n=10M$, $m=1k$ with varying $l$ and sparsity levels in $\V$.

\subsubsection{Computing platform}
To evaluate the decomposition methods on Light Field (i) an 8-core CPU (Intel Core\texttrademark i7 processor) with $12$GB of RAM is used. For computations on Light Field dataset (ii), we instanciated a cluster of $16$ m3.large nodes (machines) on Amazon EC2. Each node has 16 cores (two Intel Xeon processors) at $7.5$GB of RAM per node. The synthetic datasets are evaluated on IBM iDataPlex computing cluster which has $2304$ cores in $192$ Westmere nodes ($12$ processor cores per node) at $48$GB of RAM per node.

\subsubsection{Distributed tools}
 All \sys{}'s APIs are available to the public \cite{rankmapapi15}. The \sys{} framework's sparse matrix-based model is implemented using Eigen library to represent data in a compressed column storage (CCS) format \cite{eigenweb}. It uses MPI's standard system to distribute the data and computation and is written in C++. We have also implemented the distributed update on the factorized data on Apache Spark \cite{zaharia2010spark}.

 The \sys{} framework's sparse graph-based design is implemented using \textit{GraphLab}, a high-level graph-parallel abstraction \cite{OSDI:Gon12}. GraphLab enables vertex-update-based computations. We implemented \sys{}'s customized partitioning using Graphlab's \textit{ingress} class. The proposed architectures are mapped efficiently into GraphLab API (Section \ref{sec:graphmodel}).
Note that the GraphLab framework is designed to accelerate distributed learning for sparse graphs and thus it is not suited to process dense data until we sparsify the data using CSSD.

\subsection{Scaling of CSSD }\label{sec:sddscale}
\begin{figure}[h]
\centering
\includegraphics[width=.35\textwidth]{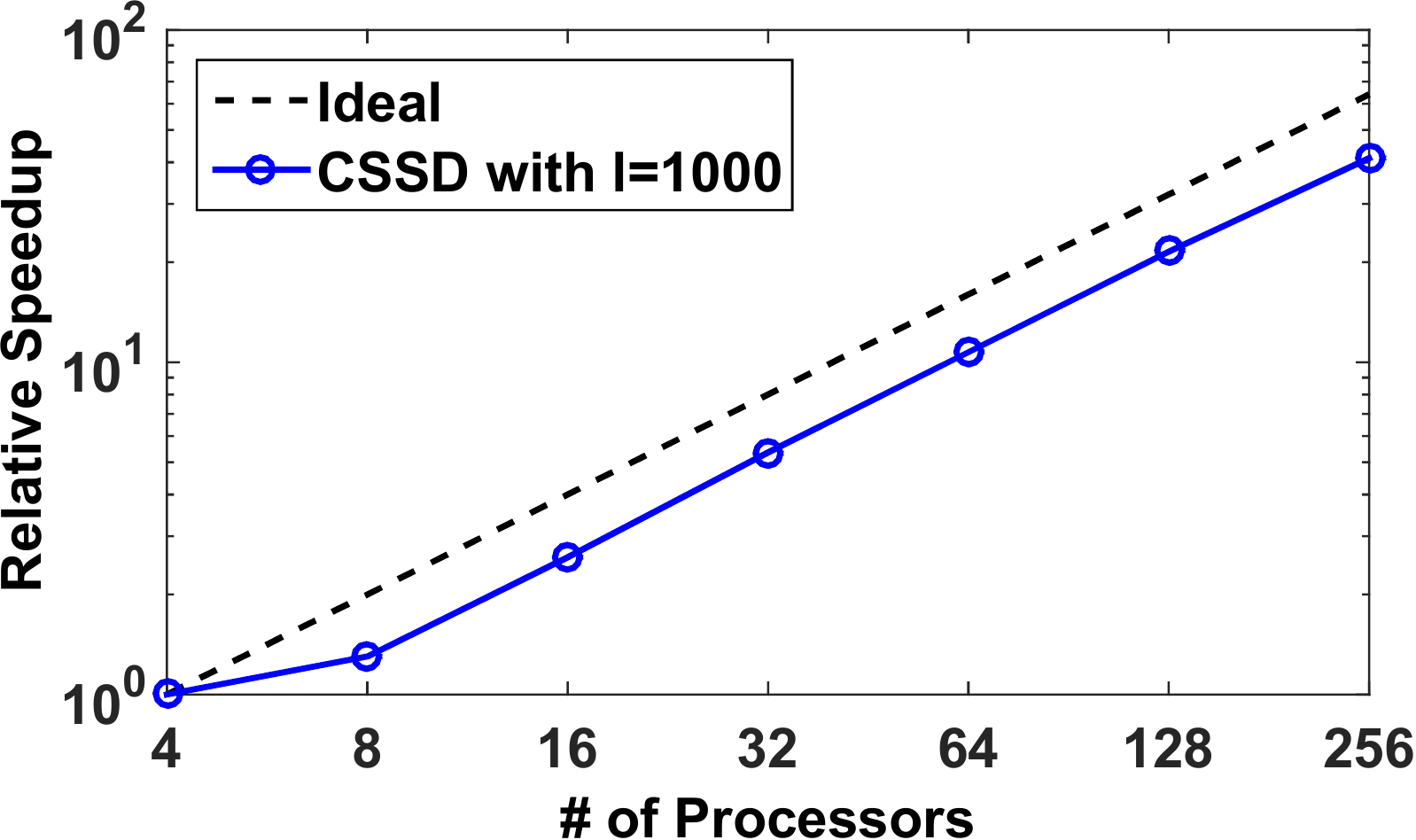}
\caption{{\em CSSD's runtime scaling behavior as the number of processors increases.}}\label{fig:vidscale}
\end{figure}

Figure \ref{fig:vidscale} shows how the runtime of CSSD scales as the number of processors increases for the VideoDict dataset. We increase the number of cores from 4 to 256 (on IBM iDataPlex cluster). The dotted line shows the ideal scale-out behavior. As can be seen, CSSD is highly parallel as it almost linearly scales with the number of processors. Thus, it can be applied to very large datasets.

\subsection{Sparse approximation}
To evaluate the performance of \sys{} for sparse approximation, we use the \textit{fast iterative shrinkage-thresholding algorithm (FISTA)} \cite{jour:Beck09} to solve the $\ell_1$-minimization problem in (\ref{eq:l1}). We study the utility of \sys{} for two applications: sparse representation-based classification for face recognition and image denoising (see Section \ref{sec:targetapps} for more details on these applications).

\subsubsection{Sparse representation-based classification for face recognition}
To employ sparse approximation for classification, our aim is to use a collection of labeled images (training set) as our dictionary $\A$ and then form a sparse representation of a test image $\y$ in terms of $\A$. After finding a sparse coefficient vector $\x$, we can then determine which signals in the testing set (columns of $\A$) are selected to represent the test signal $\y$. Based upon the class of the selected columns, we then make a decision about which class the test signal lies in. One easy way to do this is to simply sum the absolute value of the coefficients in $\x$ in a certain class and then find the class with maximum sum.

\begin{figure}[t!]
    \centering
 \includegraphics[width=.5\textwidth]{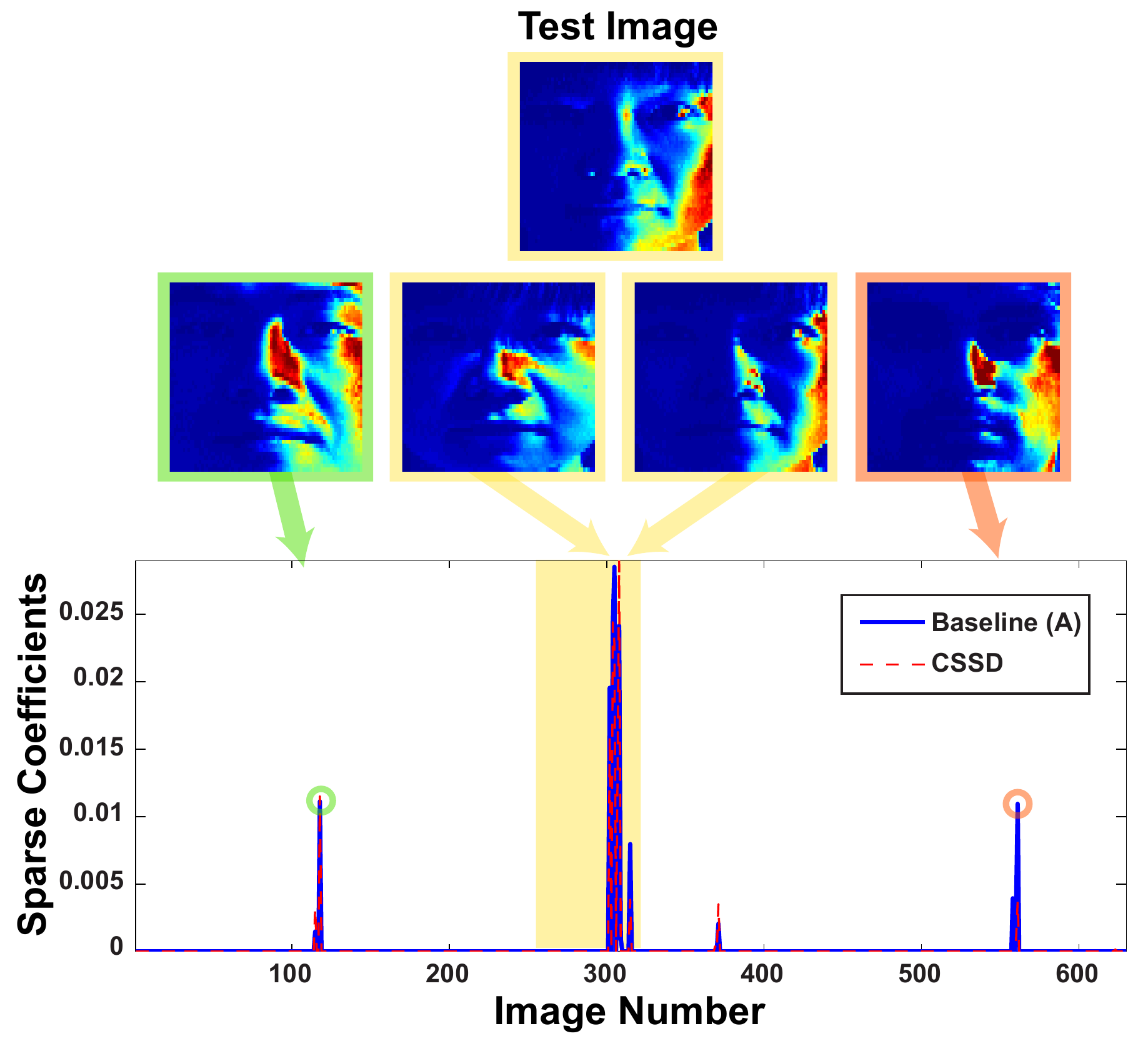}
 \caption{{\em Sparse representation-based face recognition.} We show the sparse coefficients obtained with the original Gram matrix (blue, solid) and approximate Gram matrix with CSSD for $\delta_D = 0.05$ (red, dash). On top, we show the test image, and four training images that produce significant non-zero coefficients. Both the baseline and our approach result in correct classification, as their largest coefficient is associated with a training example from the same class as the test image. The block of coefficients corresponding to images in the correct class is highlighted.}
 \label{fig:face_a}
\end{figure}

In Figure \ref{fig:face_a}, we provide a demonstration of sparse representation-based classification for face recognition. We show the test image of interest on top and the corresponding sparse coefficient vector obtained by solving (\ref{eq:l1}) with FISTA, where $\lambda = 1$. We solve FISTA with the full Gram matrix $\A^T\A$ and approximate Gram provided by CSSD, where the decomposition error $\delta_D = 0.05$ ($l = 62$).

\begin{figure*}
 \centering
 \begin{subfigure}{0.3\linewidth}
    \includegraphics[width=\textwidth]{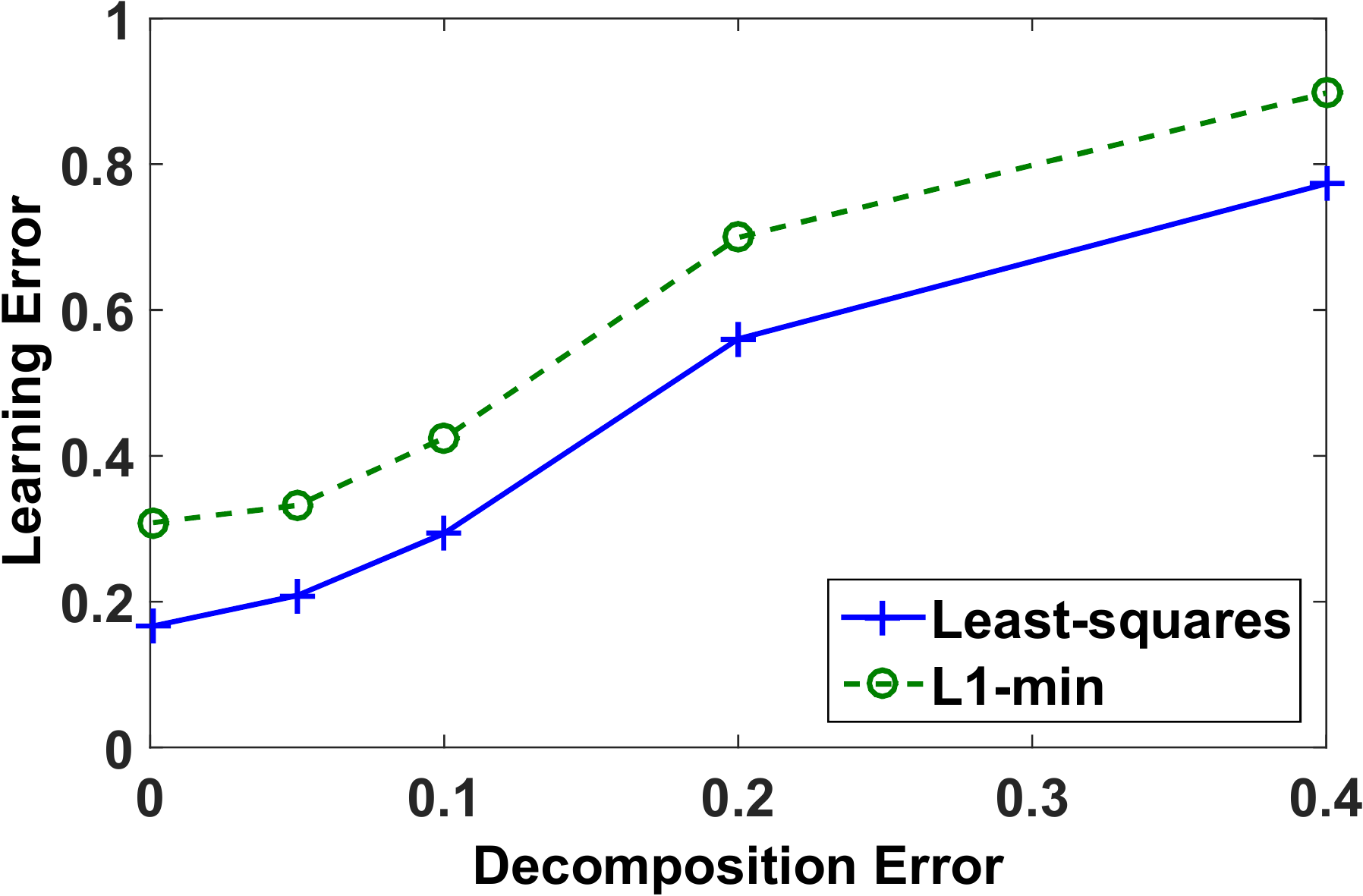}
    \caption{}\label{fig:face_c}
 \end{subfigure}
 ~
 \begin{subfigure}{0.3\linewidth}
   \includegraphics[width=\textwidth]{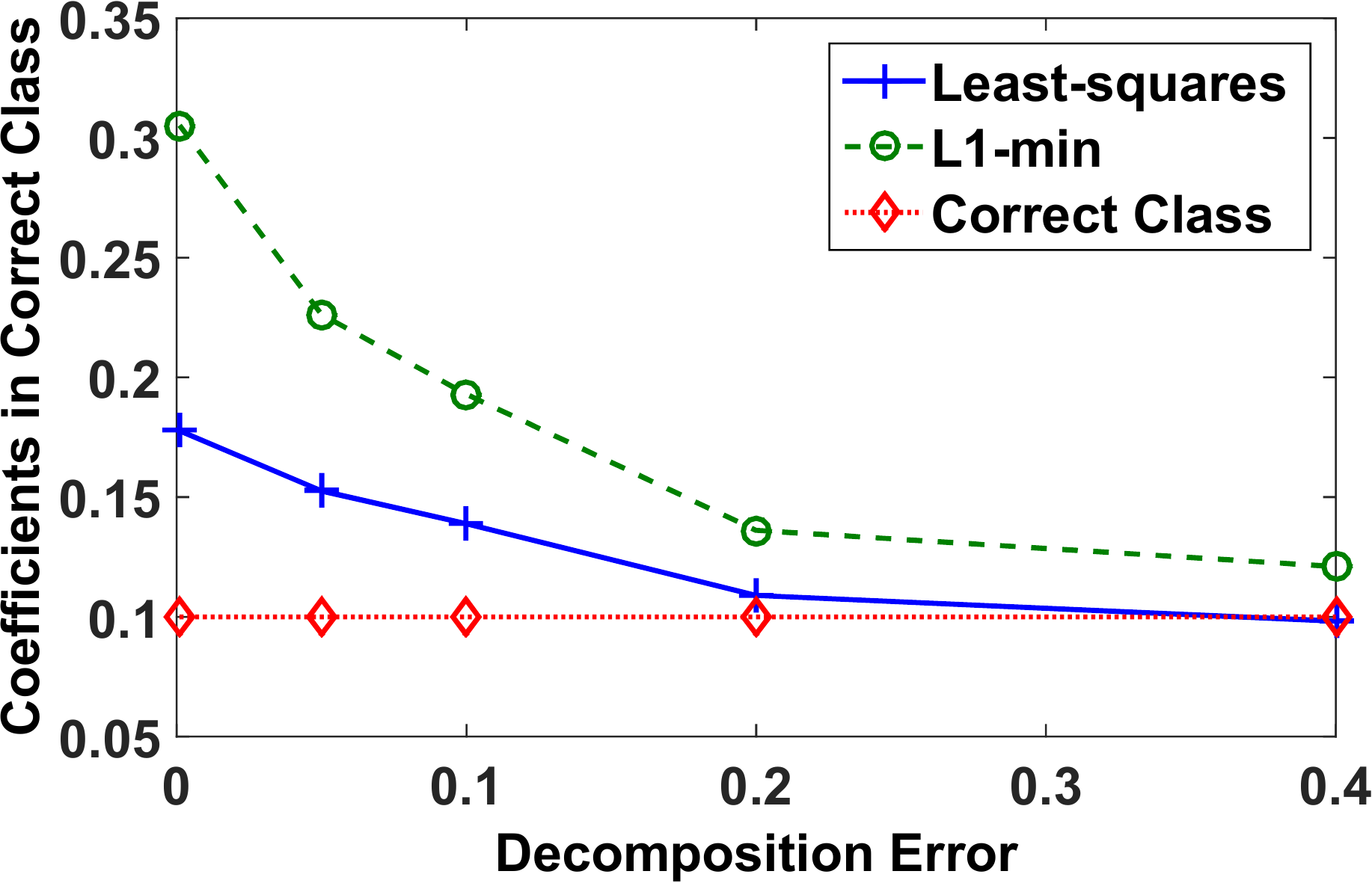}
   \caption{}\label{fig:face_d}
 \end{subfigure}
   ~
 \begin{subfigure}{0.3\linewidth}
   \includegraphics[width=\textwidth]{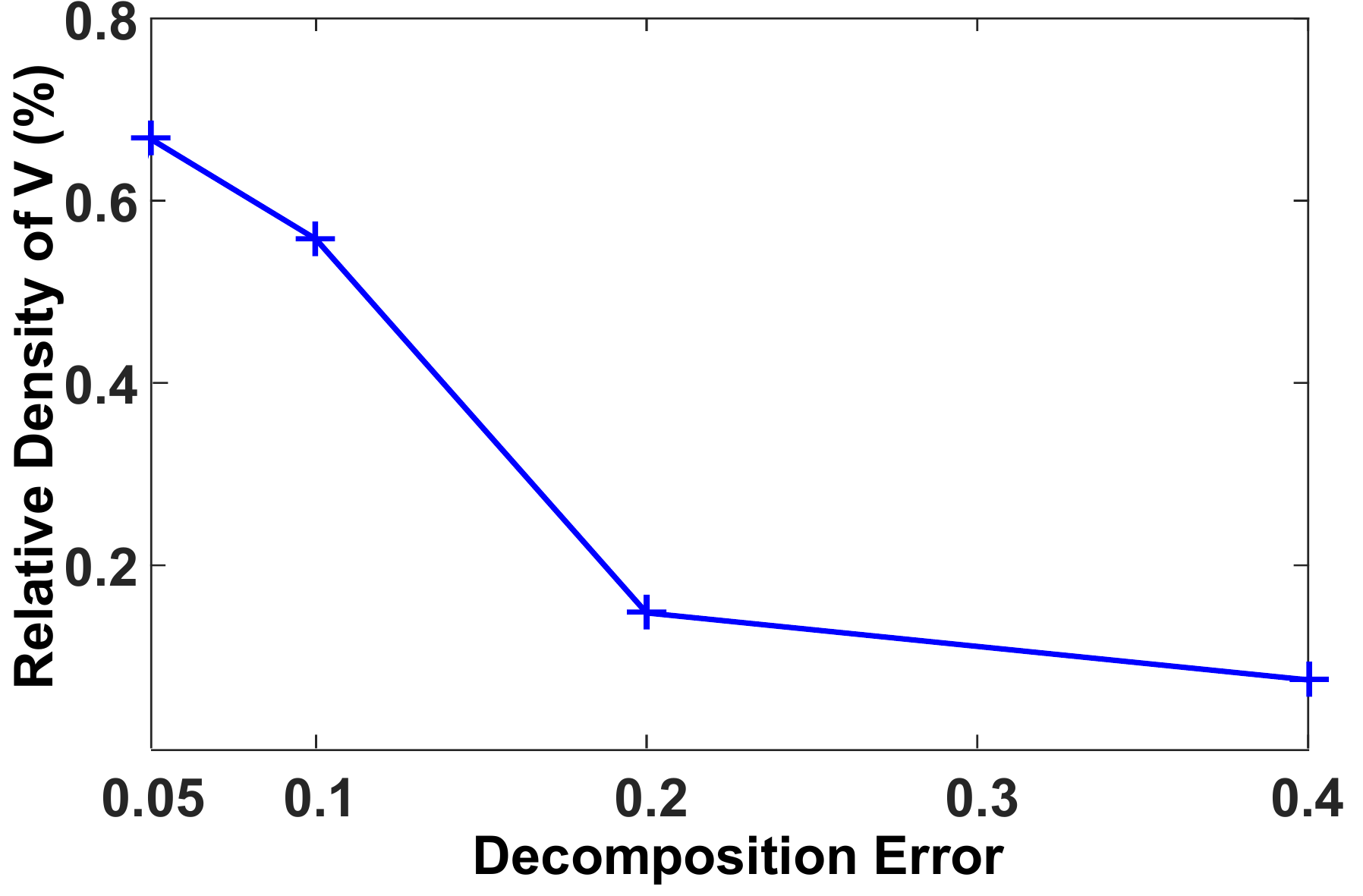}
   \caption{}\label{fig:face_b}
 \end{subfigure}
 \caption{{\em Studying the impact of the decomposition error on learning accuracy.} For a range of decompositions (varying $\delta_D$) we show (a) learning accuracy which measures the 2-norm between the solution obtained with
the full and approximate Gram, (b) sum of coefficients in the correct class, the minimum sum of coefficients required to correctly classify the test image is also shown. and (c) relative number of non-zeros in $\V$ versus the number
of non-zeros in $\A$.}\label{fig:faceclass}
\end{figure*}

To understand the connection between the decomposition error and learning accuracy for face recognition, we solve (\ref{eq:l1}) using FISTA for two different regularization parameters $\lambda = \{ 0, 1\}$, where $\lambda = 0$ corresponds to the least-squares solution and $\lambda = 1$ produces sparse solutions. We vary the decomposition error $\delta_D = \{ 0.4, 0.2, 0.1, 0.05\}$ and solve FISTA for $30$ different test images (after removing them from training) for each of these decompositions. We calculate the: learning accuracy by measuring the $\ell_2$-norm between the solution obtained with the full and approximate Gram (Figure \ref{fig:face_c}), the sum of coefficients in the correct class (Figure \ref{fig:face_d}), and the relative density of $\V$ (the number of non-zeros in $\V$ versus the number of non-zeros in $\A$) (Figure \ref{fig:face_b}). In Figure \ref{fig:face_d}, we also display the minimum sum of coefficients required to correctly classify the test image. In this case, we obtain the correct class with this approach when $\delta_D <0.2$. These results suggest that even when we allow a significant amount of decomposition error, correct classification is still possible.

\subsubsection{Light Field image denoising}
We evaluate \sys{}'s performance in denoising Light Field data. A Light Field is a multi-dimensional array of images where each image is captured from a slightly different viewpoint.
To reconstruct and denoise light fields, $\ell_1$-minimization (\ref{eq:l1}) is employed to find a sparse representation of a Light Field image with respect to an overcomplete Light Field dictionary consisting of a large number of Light Field image patches collected from many scenes \cite{jour:Marw13}. This dictionary can be used to reconstruct light field patches for the purpose of super-resolution and denoising.

We study the performance of \sys{} for reconstructing light field patches from noisy observations (image denoising). In all of our denoising experiments, Light Field (ii) is used. We first apply CSSD for decomposing the dictionary corresponding to the Light Field (ii) dataset, for two different values of $l = 240$ and $l = 1000$. The decomposition error $\delta_D$ is set to $0.1$. After decomposing the data, we then evaluate FISTA on the decomposed data with the matrix-based model. We compare \sys{}'s performance with that of a tailored distributed MPI-based model to evaluate FISTA on the original dataset ($\A$) using regular dense matrix representations. This implementation is denoted as the {\em baseline} in our evaluations.

Table \ref{table:fistapsnr} shows the total runtime of FISTA to achieve different PSNRs. The Peak Signal to Noise Ratio (PSNR) is the ratio between the maximum possible power of a signal and the power of the corrupting noise, is used to measure the performance of denoising algorithms. The PSNR is defined as $10\log_{10}(\frac{MAX}{\sqrt{MSE}})$ (dB), where $MAX$ is the maximum pixel value of the original image patch and $MSE$ is the mean square reconstruction error defined as $\frac{{{\| \y-\widehat{\y}\|^2}}}{m}$. Typically in image noise reduction applications, PSNR values of $30$ dB and higher are desired \cite{psnr:candes,psnr:boyd, ksvd}.

 In all the experiments, a batch of ten noisy patches are used as the input and the norm of the noise is set to $0.3$ times the norm of the input vector (PSNR=21.14). We observe that \sys{} can achieve the same PSNR orders of magnitude faster than the baseline implementation. For instance, if our desired PSNR is $30.0$dB, running FISTA on the decomposed data takes $13.9$s ($l=240$) and $162s$s ($l=1000$), while it takes $1050$s for the baseline. However, as expected, the baseline ($\A$) reaches higher PSNRs in comparison with those achieved from running FISTA on the decomposed data. Thus \sys{} can be used to tradeoff learning accuracy for speed.

\begin{table}[h]
\centering
\caption{{\em Runtime to reach to a specific output PSNR.} We apply FISTA to solve the denoising problem on $\A$ as well as two decompositions of $\A$ that are derived by setting $l$ to $240$ and $1000$ (in Algorithm 1). The lower dimensional decomposition ($l=240$) reaches to up to $30$ dB output PSNR much faster, due to its lower memory footprint and computing requirements. Similarly, $l=1000$ reaches to up to $35$ dB PSNR much faster than $\A$ but cannot reach to $40$ dB PSNR.}
\label{table:fistapsnr}
\begin{tabular}{|c||r|r|r|}
\hline
PSNR (dB) & $l=240$ & $l=1000$ & baseline ($\A$)\\ \hline \hline
25 & 4.2 & 72 & 487 \\
30 & 13.9 & 162 & 1050 \\
35 & - & 356 & 2051 \\
40 & - & - & 3171 \\
\hline
\end{tabular}
\end{table}

\subsection{Power method}
\begin{figure*}[!ht]
\centering
 \begin{subfigure}[b]{0.30\textwidth}
   \includegraphics[width=\textwidth]{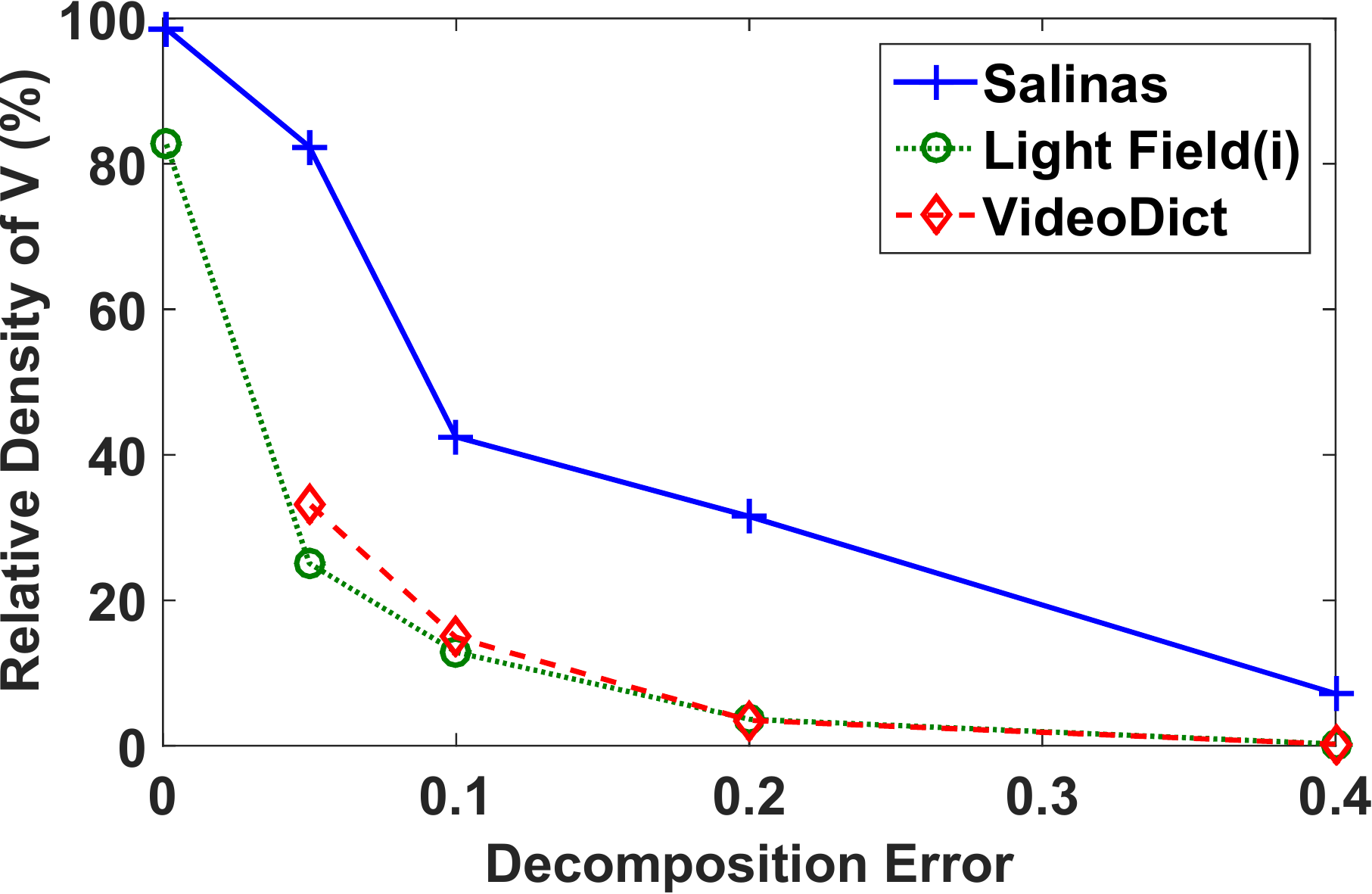}
   \caption{}\label{fig:powernnz}
 \end{subfigure}
 ~
 \begin{subfigure}[b]{0.30\textwidth}
 \includegraphics[width=\textwidth]{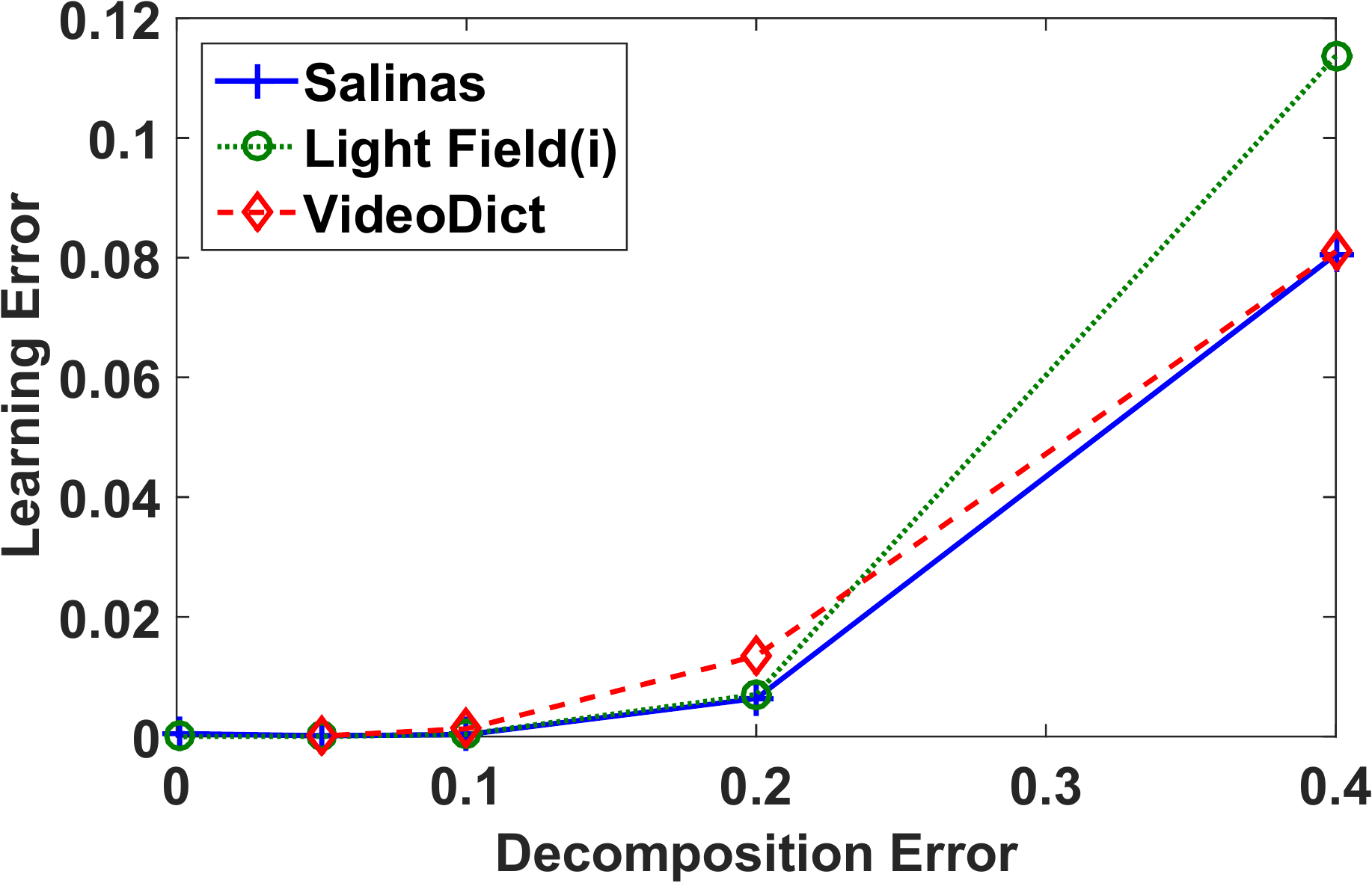}
 \caption{}\label{fig:powerdeltal}
 \end{subfigure}
 ~
 \begin{subfigure}[b]{0.30\textwidth}
 \includegraphics[width=\textwidth]{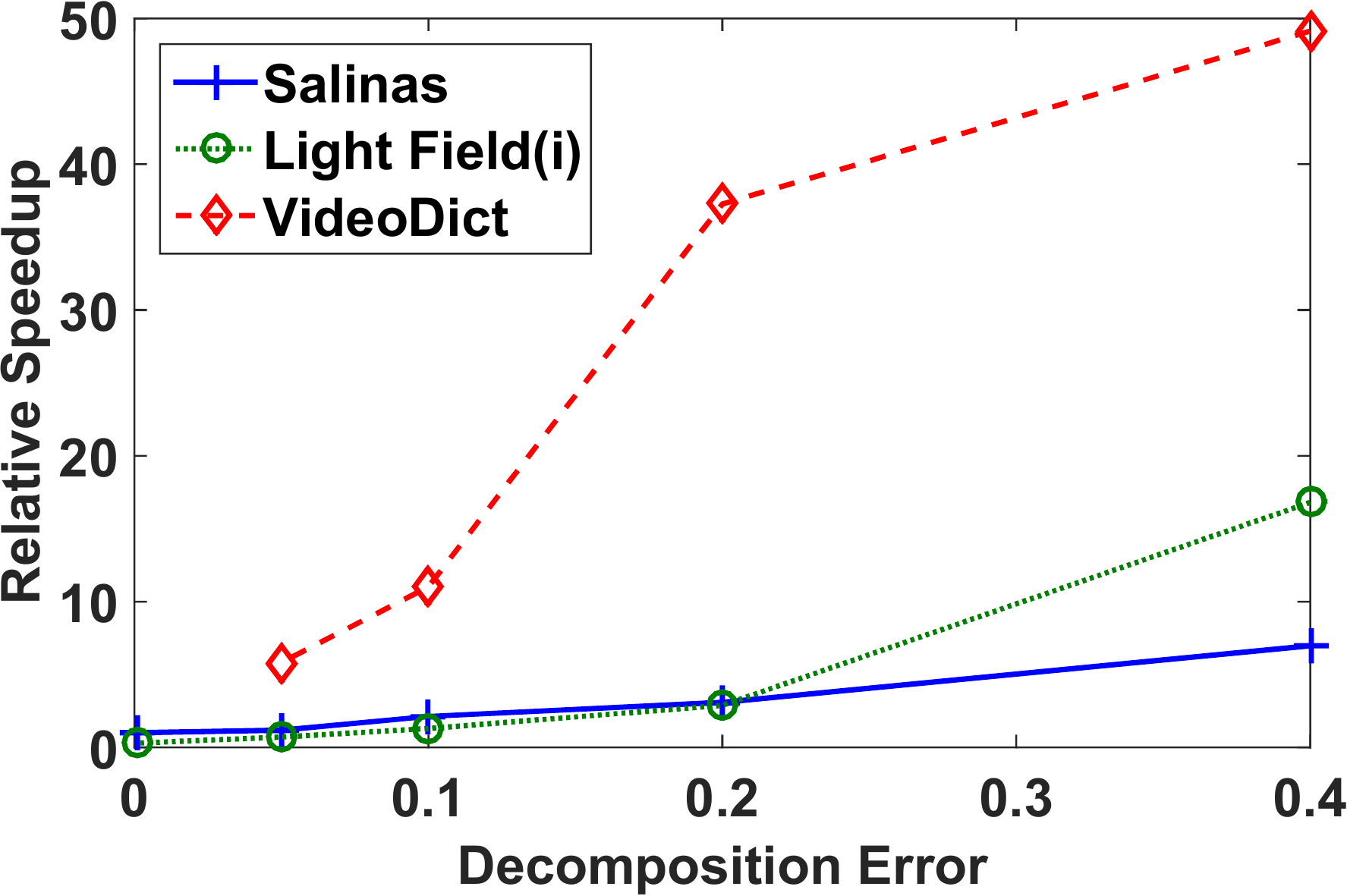}
 \caption{}\label{fig:powertime}
 \end{subfigure}
 \caption{ {\em Power method results for three datasets: Salinas, VideoDict, and Light Field (i).} In (a), we show the effect of varying the decomposition error ($\delta_D$) on the sparsity of the factor $\V$. Reported values are the number of non-zeros in $\V$ normalized to the number of non-zeros in $\A$. In (b), we show the learning error ($\delta_L$) and in (c) the relative speedup of power method to find the first 100 eigenvalues. Relative values are runtime of power method using the decomposed factors derived from CSSD normalized to the corresponding runtime using $\A$.}
\label{fig:powermethod}
\end{figure*}

We also evaluate our framework on power method for three datasets: Salinas, VideoDict, and Light Field (i) (see Section \ref{sec:targetapps} for more discussion of the power method). The matrix-based model is used and the experiments are run on 64 cores on an IBMiDataplex cluster. We run CSSD with various decomposition errors ($\delta_D$) that belong to the following set: $\{ 0.4, 0.2, 0.1, 0.05, 0.001\}$ and run the power method on each of the decomposition results. Figure \ref{fig:powernnz} shows the sparsity of $\V$ as we vary the error. As expected, for larger error tolerances, a sparser decomposition is achieved. Figure \ref{fig:powerdeltal} shows the impact of the decomposition error on the accuracy of the results of the power method. Here, the learning error ($\delta_L$) is defined to be the normalized accumulated error of the first 100 eigenvalues. By lowering the decomposition error, we can observe significant improvements in the accuracy of the power method. Finally, Figure \ref{fig:powertime} shows the corresponding normalized runtimes to find the first 100 eigenvalues. Our results suggest that significant speedups are achieved in comparison with the baseline.

\subsection{Graph- vs. matrix-based models} \label{ssec:evalmodel}
We compare the performance of \sys{}'s vertex and matrix-based models for various synthetic decomposed data. The purpose of these evaluations is to determine the advantages of each of the model, with respect to the structure of the data. In all the experiments, the iterative update in (\ref{eq:update}) is applied on a random input vector $\y$. The experiments are done on an IBM iDataPlex computing cluster. In all the figures, the runtime results for the dense matrix-based implementation (i.e., regular deployment of the decomposed matrices without using CCS format) are provided to demonstrate the efficiency achieved by exploiting sparsity in $\V$ through sparse matrix-based and graph-based models. For the former model, we report analysis based on our C++ MPI implementation and for the latter model we report analysis based on our modification of GraphLab engine to implement \sys{}.

\begin{figure*}
 \centering
 \begin{subfigure}[b]{0.30\textwidth}
    \includegraphics[width=\textwidth]{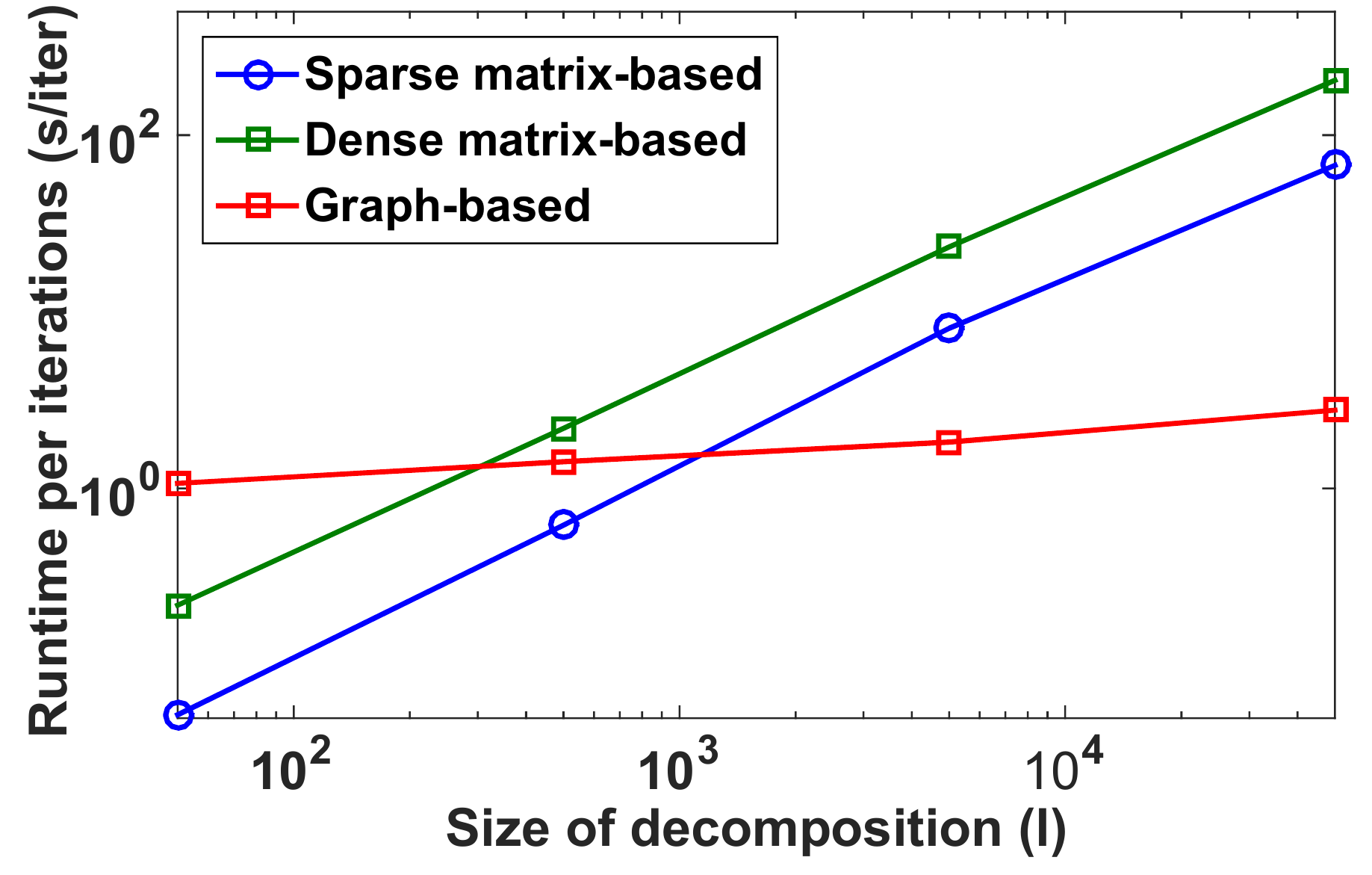}
    \caption{}\label{fig:figure_e3}
 \end{subfigure}
 ~
 \begin{subfigure}[b]{0.30\textwidth}
   \includegraphics[width=\textwidth]{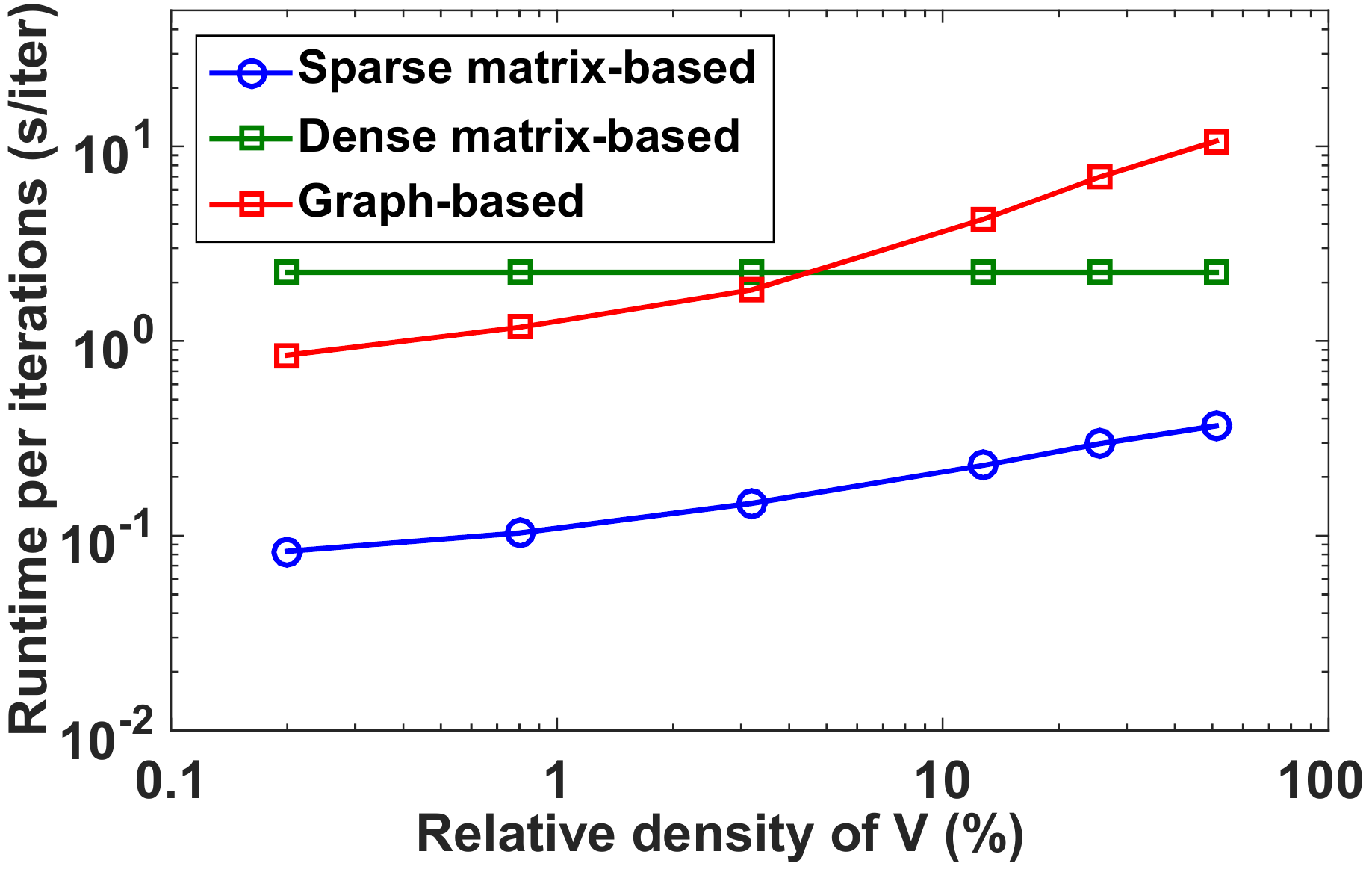}
   \caption{}\label{fig:figure_e1}
 \end{subfigure}
 ~
 \begin{subfigure}[b]{0.30\textwidth}
   \includegraphics[width=\textwidth]{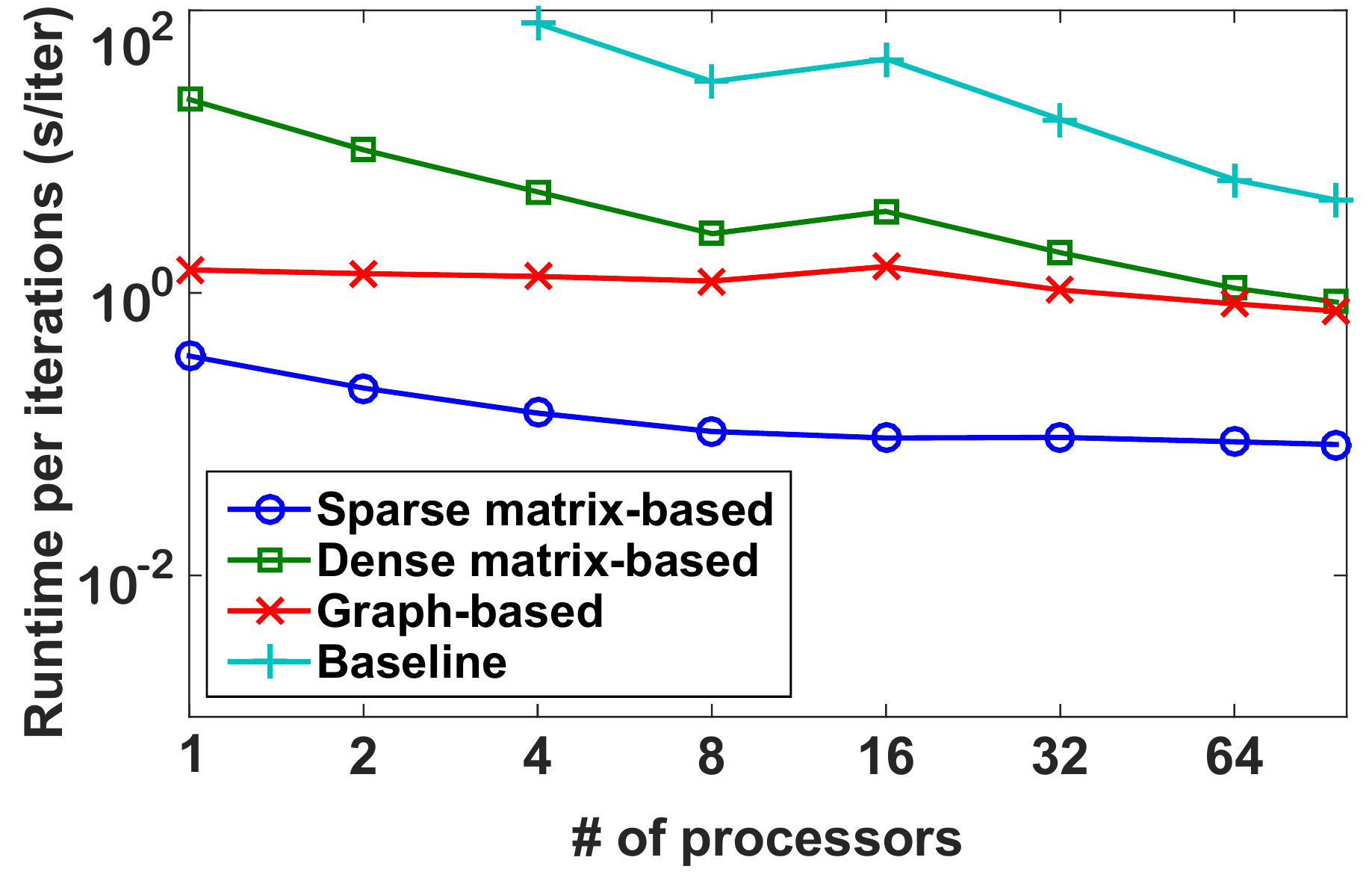}
   \caption{}\label{fig:timescaled}
 \end{subfigure}
 \caption{ {\em Comparison of matrix- and graph-based computational models on synthetic block diagonal data.} We compare the (a) runtime vs. size of factorization, (b) relative density of $\V$, and (c) number of processors.}
\end{figure*}

The performance of \sys{} for different (block-diagonal) $\V$'s, with fixed number of non-zeros (set to $100M$), is shown in Figure \ref{fig:figure_e3}. As $l$ increases, the density-level of $\V$ decreases. The graph-based model's performance is more consistent as $l$ increases. However, the matrix-based model's performance degrades for larger $l$'s. This observation can also be explained due to the fact that the communication overhead of the matrix-based model is more affected by larger $l$'s.

Figure \ref{fig:figure_e1} shows the performance for a fixed $l=500$ on block-diagonal matrices $\V$, for varying densities of $\V$. As density increases, the performance decreases in both models. However, the performance degradation in graph-based model is worse due to the overhead of representing a large number of edges. Lastly, Figure \ref{fig:timescaled} shows the scaling performance of the models for various number of processors. When the number of processors is less than $12$, the computations are done on a single node. Thus the reverse scaling behavior while increasing the number of processors from $8$ to $16$ is due to the high overhead of the inter-node communication cost. For comparison purposes, we report the scaling of the baseline approach that operates on the original dense $m=1k$ by $n=10M$ dataset, instead of its decomposed form. It can be seen that as the number of processors increases, the performance gap between different methods shrink. However, even with a large number of processors ($\ge 100$), the decomposed models perform up to $2$ orders of magnitude better than the baseline.

These experiments provide insight into the use-case of each model. Depending on the structure of the decomposed data and the specifications of the platform, an appropriate model should be selected. A more systematic domain-specific approach for model selection is the subject of future work.

\subsection{Memory Analysis} Table \ref{tab:memory} compares the required memory for storing matrices $\V$ and $\D$. The memory usage of the original matrix $\A$ is also provided. We also provided the memory savings for the case in which $\D$ is formed in the same fashion as CSSD but $\V$ is computed via least-squares, as opposed to OMP. \sys{} results in up to 77.8$\times$ (memory usage) improvement over $\A^T\A$ and 8.6$\times$ improvement over the adaptive norm-2 projection based decomposition. The approximation error for both decomposition methods is set to $\delta_D=0.1$.

\begin{table}[H]
\centering
\scalebox{.99}{
\begin{tabular}{|c|c|c|c|}
\hline
 &  &  &  \\
Memory size & Original & Least-squares & RankMap   \\
(MB)        &  data    &      &                \\\hline
Salina & 87.9  & 86.9  & 38.2
\\ \hline
VideoDict & 1411.2 &835.0 & 279.2
\\ \hline
Light Field (ii) &  14796.8 & 1634.8  &  190.1
 \\ \hline
\end{tabular}
}\caption{Memory analysis. RankMap achieves significant improvement via its adaptive and sparsity-inducing approach.}\label{tab:memory}
\end{table}

\subsection{Comparison with Spark}
We have already shown the results based on our implementation of \sys{} on GraphLab in Figures \ref{fig:figure_e3}, \ref{fig:figure_e1}, and \ref{fig:timescaled}. Now we provide runtime comparisons between \sys{} and a Apache Spark-based implementation of the power method on the baseline $\A$ versus on the decomposed data $\D\V$. Recall that the core iterative update function used in power method is provided in (\ref{eq:updatepower}). We report the average runtime per iteration with Spark and our implementations on the same hardware.

\begin{figure}
        \centering
	\includegraphics[width=\columnwidth]{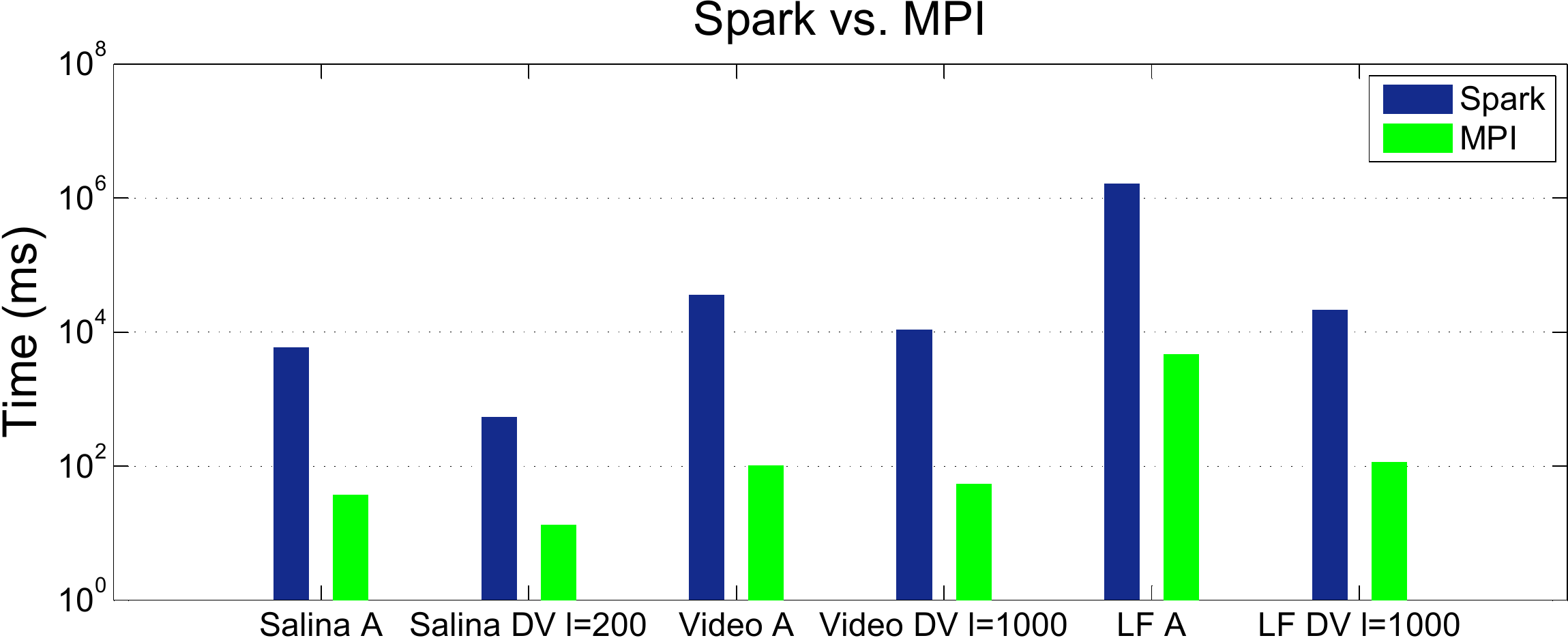}
	\caption{{\em The average runtime per power method iteration for RankMap and a Spark implementation of the baseline method.}}
	\label{fig:sparkvsMPI}
	\vspace{-4mm}
\end{figure}

Figure \ref{fig:sparkvsMPI} shows the average runtime per iteration on a cluster of 8 nodes, 8 core per processor for Salinas, VideoDict, and Light Field (ii) datasets. As expected, our carefully tailored implementation of \sys{} based on C++ MPI, performed significantly better than Spark, by more than 2 orders of magnitude in some cases. This gap in performance is in part due to our particular implementation of \sys{}, which carefully partitions the data such that the computation per core is balanced and the communication is reduced (see Section 6.3.2 for performance bounds). In addition, Spark provides fault-tolerance which causes extra overhead due to data replications.

\section{Discussion}\label{sec:ext}
This paper introduced \sys{}, a novel distributed framework for applying a host of iterative learning algorithms on large-scale dense and structured datasets. Our framework leverages {\em low-dimensional structure} in datasets in order to quickly map a large dataset with dense dependencies into lower dimensional components with sparse representations. We introduce two computational models, a matrix- and graph-based model, that can be used to execute distributed learning algorithms. Our framework provides an efficient partitioning of the computational flow that guarantees load balancing and significantly lowers communication overhead. We apply our matrix- and graph-based models to numerous real-world and synthetic datasets and demonstrate significant improvements in the runtime and memory footprint.

There is an unavoidable cost associated with factorizing the data. For extremely large datasets however, this initial cost can pay off a lot. The performance gain achieved on the subsequent computations justifies the one-time decomposition overhead. For example, we decompose Light Field (ii) dataset (Section \ref{sec:eval}) on a cluster of $4$ nodes (each with 12 cores) on IBM iDataPlex. For $l=240$, the decomposition is completed in less than $15$ minutes. For $10$ sample patches (each of length $18k$) the overall reconstruction time is reduced from more than $1000$s to below $20s$. Thus, the offline decomposition overhead can be justified once considering that there are thousands of patches in a single light field. Moreover, the same dictionary can be used to reconstruct other light field datasets.

There are a number of existing column sampling-based methods that aim to improve the performance of specific learning objectives, such as least-squares \cite{LSRN14}, $\ell_2$-minimization with square root $\ell_1$ penalty \cite{squarelasso14}, and SVM \cite{CSS_SVM_Fine01}.   \sys{} is unique in that uses column sampling to improve the performance of a broad class of ML algorithms that operates on the Gram matrix. Moreover,  \sys{} relies of the sparsity of the decomposition for further improvement in runtime, energy and memory usage. Finally, to the best of our knowledge \sys{} is the first end-to-end framework that is equipped with open-source supported APIs \cite{rankmapapi15}. 

Sparse matrix factorization approaches such as SPCA and KSVD have objectives similar to CSSD, however, their complexity make them difficult to apply to massive datasets. As we sample the dataset instead of learning a factorization of the data, our proposed decomposition is faster and scalable. Whilst our sampling-based approach is effective, the decomposition phase in \sys{} (see Figure \ref{fig:sys}) can be readily replaced by other sparse decomposition approaches. Tradeoffs between the time to compute a factorization (via learning or sampling) and how sparse we can make the decomposition are likely to exist. Although outside of the scope of this current work, it would be interesting to study the utility of learning in terms of its later computational benefit.

Our graph-based and matrix-based computational models provide advantages in different data regimes. Thus, it is natural to ask which model to select for data processing. Both models follow the same computational flow and operate only on the non-zeros. In practice, we observe that the matrix-based approach is faster: this is especially true when we exploit sparsity in the decomposition with a sparse matrix-based approach. This is likely due to the fact that the graph-based model requires extra overhead to store and operate on the vertices and edges. The main advantage of the graph-based model is in its reduced communication cost (Section \ref{ssec:graphper}). When $\V$ is completely block diagonal the communication becomes almost independent of the number of computing nodes $n_c$ and is only proportional to $2l$. In contrast, the communication cost in the matrix-based model is always proportional to $2ln_c$. This difference may result in a better overall performance of the graph-based approach, especially for larger $l$ and $n_c$ values. In general, the performance of each model is highly dependent on the specifications of the available computing nodes including the communication bandwidth and computation power (e.g., maximum floating point operations per second). Our evaluations in Section \ref{ssec:evalmodel} provide further insight into the differences of the two models.

Throughout our experiments, we used FISTA, an accelerated gradient descent method, as an optimizer. Our computation/communication and memory minimizing framework can also be applied to other optimization methods such as Stochastic Gradient Descent (SGD) \cite{recht2011hogwild} and Stochastic Coordinate Descent (SCD) \cite{coord:Liu:2015}. Both SGD and SCD operate on the entire $m\times n$ dataset, however, each iterative update is performed on a subset of rows (as in SGD) or along the columns (in SCD). For this reason, the convergence of stochastic method is slower. SGD can be integrated within \sys{} by breaking the $m$ rows of matrix $\A$ into batches, and performing \sys{}'s decomposition on each batch.  SCD can also be applied to the factorized dataset $\D\V$ instead of $\A$. In general, \sys{} is not limited to a particular optimizer, it is beneficial whenever there is a need to store/ and or iteratively perform matrix multiplication on large datasets.

In this work, we show how sparse matrix factorization and adaptive sampling can be used to speed up iterative optimization algorithms on large datasets. We have mainly explored its use for computational gains, however, recent theoretical results have shown that subsampling data can also be beneficial for learning \cite{rudi2015less}. \sys{} provides a new computational framework from which we can begin to test the ideas of efficiency, both in terms of quality of learning and computing performance, through randomization and subsampling.

\vspace{-2mm}

\ifsubmission
  \bibliographystyle{plain}
\else
  \bibliographystyle{alpha} 
\fi
\bibliography{main_bib}

\begin{thebibliography}{10}

\bibitem{salina}
Hyperspectral remote sensing scenes.
\newblock
  \url{http://www.ehu.eus/ccwintco/index.php?title=Hyperspectral_Remote_Sensing_Scenes}.

\bibitem{lfweb}
The light field archive.
\newblock \url{http://lightfield.stanford.edu/}.

\bibitem{rankmapapi15}
Rankmap {APIs}.
\newblock \url{https://github.com/azalia/RankMap}.

\bibitem{ksvd}
M~Aharon, M~Elad, and A~Bruckstein.
\newblock {SVD}: An algorithm for designing overcomplete dictionaries for
  sparse representation.
\newblock {\em IEEE Trans Sig. Process.}, 54(11):4311--4322, 2006.

\bibitem{modelcs}
R~G Baraniuk, V~Cevher, M~F Duarte, and C~Hegde.
\newblock Model-based compressive sensing.
\newblock {\em IEEE Trans. Inf. Theory}, 56(4):1982--2001, 2010.

\bibitem{jour:Beck09}
A~Beck and M~Teboulle.
\newblock A fast iterative shrinkage-thresholding algorithm for linear inverse
  problems.
\newblock {\em SIIMS}, 2(1):183--202, 2009.

\bibitem{psnr:boyd}
E~J Candes, M~B Wakin, and S~P Boyd.
\newblock Enhancing sparsity by reweighted $\ell_1$ minimization.
\newblock {\em JFAA}, 14(5-6):877--905, 2008.

\bibitem{jour:chen1998}
K~Chen.
\newblock On a class of preconditioning methods for dense linear systems from
  boundary elements.
\newblock {\em SISC}, 20(2):684--698, 1998.

\bibitem{chen1998atomic}
S~S Chen, D~L Donoho, and M~A Saunders.
\newblock Atomic decomposition by basis pursuit.
\newblock {\em SISC}, 20(1):33--61, 1998.

\bibitem{cortes2010impact}
C~Cortes, M~Mohri, and A~Talwalkar.
\newblock On the impact of kernel approximation on learning accuracy.
\newblock In {\em TAISTATS}, pages 113--120, 2010.

\bibitem{daubechies2004iterative}
I~Daubechies, M~Defrise, and C~De~Mol.
\newblock An iterative thresholding algorithm for linear inverse problems with
  a sparsity constraint.
\newblock {\em Comm. Pure Appl. Math.}, pages 1413--1457, 2004.

\bibitem{davisOMP}
G~M Davis, S~G Mallat, and Z~Zhang.
\newblock Adaptive time-frequency decompositions.
\newblock {\em OE}, 33(7):2183--2191, 1994.

\bibitem{jour:dean2008}
J~Dean and S~Ghemawat.
\newblock {MapReduce}: simplified data processing on large clusters.
\newblock {\em CACM}, 51(1):107--113, 2008.

\bibitem{deshpande2006matrix}
A~Deshpande, L~Rademacher, S~Vempala, and G~Wang.
\newblock Matrix approximation and projective clustering via volume sampling.
\newblock In {\em SODA}, pages 1117--1126. SIAM, 2006.

\bibitem{jour:Drin04}
P~Drineas, A~Frieze, R~Kannan, S~Vempala, and V~Vinay.
\newblock Clustering large graphs via the singular value decomposition.
\newblock {\em Machine learning}, 56(1-3):9--33, 2004.

\bibitem{drineas2005nystrom}
P~Drineas and M~W Mahoney.
\newblock On the {Nystr{\"o}m} method for approximating a gram matrix for
  improved kernel-based learning.
\newblock {\em JMLR}, 6:2153--2175, 2005.

\bibitem{jour:dyerarxiv15}
E~L Dyer, T~A Goldstein, R~Patel, K~P Kording, and R~G Baraniuk.
\newblock Self-expressive decompositions for matrix approximation and
  clustering.
\newblock {\em arXiv:1505.00824}, 2015.

\bibitem{DyerJMLR13}
E~L Dyer, A~C Sankaranarayanan, and R~G Baraniuk.
\newblock Greedy feature selection for subspace clustering.
\newblock {\em JMLR}, 14(1):2487--2517, 2013.

\bibitem{vidaljournal}
E~Elhamifar and R~Vidal.
\newblock Sparse subspace clustering: Algorithm, theory, and applications.
\newblock {\em TPAMI}, 35(11):2765--2781, 2013.

\bibitem{CSS_SVM_Fine01}
S~Fine and K~Scheinberg.
\newblock Efficient {SVM} training using low-rank kernel representations.
\newblock {\em JMLR}, 2(Dec):243--264, 2001.

\bibitem{fowlkes2004spectral}
C~Fowlkes, S~Belongie, F~Chung, and J~Malik.
\newblock Spectral grouping using the {Nystr{\"o}m} method.
\newblock {\em TPAMI}, 26(2):214--225, 2004.

\bibitem{yaleb}
A~S Georghiades, P~N Belhumeur, and D~J Kriegman.
\newblock From few to many: Illumination cone models for face recognition under
  variable lighting and pose.
\newblock {\em TPAMI}, 23(6):643--660, 2001.

\bibitem{systemml11}
A~Ghoting, R~Krishnamurthy, E~Pednault, B~Reinwald, V~Sindhwani, S~Tatikonda,
  Y~Tian, and S~Vaithyanathan.
\newblock {SystemML}: Declarative machine learning on {MapReduce}.
\newblock In {\em ICDE}, pages 231--242. IEEE, 2011.

\bibitem{ICML:Git13}
A~Gittens and M~W Mahoney.
\newblock Revisiting the nystrom method for improved large-scale machine
  learning.
\newblock {\em ICML}, pages 567--575, 2013.

\bibitem{OSDI:Gon12}
J~E Gonzalez, Y~Low, H~Gu, D~Bickson, and C~Guestrin.
\newblock {PowerGraph}: Distributed graph-parallel computation on natural
  graphs.
\newblock In {\em OSDI}, pages 17--30, 2012.

\bibitem{jour:gray2000}
A~G Gray and A~W Moore.
\newblock {`N-Body'} problems in statistical learning.
\newblock In {\em NIPS}, pages 521--527. MIT Press, 2001.

\bibitem{eigenweb}
Ga{\"e}l Guennebaud, Benoit Jacob, et~al.
\newblock \url{http://eigen.tuxfamily.org/}.

\bibitem{hitomi2011video}
Y~Hitomi, J~Gu, M~Gupta, T~Mitsunaga, and S~Nayar.
\newblock Video from a single coded exposure photograph using a learned
  over-complete dictionary.
\newblock In {\em ICCV}, pages 287--294. IEEE, 2011.

\bibitem{hoerl1970ridge}
A~E Hoerl and R~W Kennard.
\newblock Ridge regression: Biased estimation for nonorthogonal problems.
\newblock {\em Technometrics}, 12(1):55--67, 1970.

\bibitem{journee2010generalized}
M~Journ{\'e}e, Y~Nesterov, P~Richt{\'a}rik, and R~Sepulchre.
\newblock Generalized power method for sparse principal component analysis.
\newblock {\em JMLR}, 11(Feb):517--553, 2010.

\bibitem{kanatani01}
K~Kanatani.
\newblock Motion segmentation by subspace separation and model selection.
\newblock In {\em ICCV}, volume~2, pages 586--591, 2001.

\bibitem{Lin:2007:pgd}
C~Lin.
\newblock Projected gradient methods for nonnegative matrix factorization.
\newblock {\em Neural Comput.}, 19(10):2756--2779, 2007.

\bibitem{coord:Liu:2015}
J~Liu, S~J Wright, C~R{\'e}, V~Bittorf, and S~Sridhar.
\newblock An asynchronous parallel stochastic coordinate descent algorithm.
\newblock {\em JMLR}, pages 285--322, 2015.

\bibitem{jour:low2010}
Y~Low, J~E Gonzalez, A~Kyrola, D~Bickson, C~E Guestrin, and J~Hellerstein.
\newblock {GraphLab}: A new parallel framework for machine learning.
\newblock {\em UAI}, pages 340--349, 2010.

\bibitem{jour:malewicz2010}
G~Malewicz, M~H Austern, A~JC Bik, J~C Dehnert, I~Horn, N~Leiser, and
  G~Czajkowski.
\newblock {Pregel}: a system for large-scale graph processing.
\newblock In {\em SIGMOD}, pages 135--146. ACM, 2010.

\bibitem{jour:Marw13}
K~Marwah, G~Wetzstein, Y~Bando, and R~Raskar.
\newblock Compressive light field photography using overcomplete dictionaries
  and optimized projections.
\newblock {\em TOG}, 32(4):46, 2013.

\bibitem{LSRN14}
X~Meng, M~A Saunders, and M~W Mahoney.
\newblock {LSRN}: A parallel iterative solver for strongly over-or
  underdetermined systems.
\newblock {\em SISC}, 36(2):95--118, 2014.

\bibitem{book:montgomery2012}
D~C Montgomery, E~A Peck, and G~G Vining.
\newblock {\em Introduction to linear regression analysis}.
\newblock John Wiley \& Sons, 2015.

\bibitem{conf:nodelman2012}
U~Nodelman, C~R Shelton, and D~Koller.
\newblock Expectation maximization and complex duration distributions for
  continuous time bayesian networks.
\newblock {\em arXiv preprint arXiv:1207.1402}, 2012.

\bibitem{squarelasso14}
V~Pham and L~El Ghaoui.
\newblock Robust sketching for multiple square-root {LASSO} problems.
\newblock In {\em AISTATS}, 2015.

\bibitem{facesubs}
R~Ramamoorthi.
\newblock Analytic {PCA} construction for theoretical analysis of lighting
  variability in images of a lambertian object.
\newblock {\em TPAMI}, 24(10):1322--1333, 2002.

\bibitem{recht2011hogwild}
B~Recht, C~Re, S~Wright, and F~Niu.
\newblock {HOGWILD}: A lock-free approach to parallelizing stochastic gradient
  descent.
\newblock In {\em NIPS}, pages 693--701, 2011.

\bibitem{rubinstein2008efficient}
R~Rubinstein, M~Zibulevsky, and M~Elad.
\newblock Efficient implementation of the {K-SVD} algorithm using batch
  orthogonal matching pursuit.
\newblock {\em Technion, Tech. Report}, 40(8):1--15, 2008.

\bibitem{rudi2015less}
A~Rudi, R~Camoriano, and L~Rosasco.
\newblock Less is more: {Nystr{\"o}m} computational regularization.
\newblock In {\em NIPS}, pages 1657--1665, 2015.

\bibitem{psnr:candes}
J~Starck, E~J Cand{\`e}s, and D~L Donoho.
\newblock The curvelet transform for image denoising.
\newblock {\em IEEE Trans Image Processing}, 11(6):670--684, 2002.

\bibitem{thompson72}
R~C Thompson.
\newblock Principal submatrices {IX}: Interlacing inequalities for singular
  values of submatrices.
\newblock {\em Linear Algebra and its Applications}, 5(1):1--12, 1972.

\bibitem{wright2010sparse}
J~Wright, Y~Ma, J~Mairal, G~Sapiro, T~S Huang, and S~Yan.
\newblock Sparse representation for computer vision and pattern recognition.
\newblock {\em Proceedings of the IEEE}, 98(6):1031--1044, 2010.

\bibitem{conf:yedidia2000}
J~S Yedidia, W~T Freeman, Y~Weiss, et~al.
\newblock Generalized belief propagation.
\newblock {\em NIPS}, pages 689--695, 2000.

\bibitem{zaharia2010spark}
M~Zaharia, M~Chowdhury, T~Das, A~Dave, J~Ma, M~McCauley, M~J Franklin,
  S~Shenker, and I~Stoica.
\newblock Resilient distributed datasets: A fault-tolerant abstraction for
  in-memory cluster computing.
\newblock In {\em NSDI}, pages 2--2. USENIX, 2012.

\bibitem{conf:zaharia2010}
M~Zaharia, M~Chowdhury, M~J Franklin, S~Shenker, and I~Stoica.
\newblock {Spark}: Cluster computing with working sets.
\newblock In {\em HotCloud}, pages 10--10. USENIX, 2010.

\bibitem{jour:zou2006}
H~Zou, T~Hastie, and R~Tibshirani.
\newblock Sparse principal component analysis.
\newblock {\em J. Comp. Graph. Stat.}, 15(2):265--286, 2006.

\end{thebibliography}

\end{document}